**Highlights**

- A linked data approach connects IoT data with BIM

- Acquisition and ingestion, Batch analytics and Integration supports the overall integration

- Time-series navigation of summarized data is facilitated with visual programming languages

- Lab- and full-scale case studies demonstrate the scalability of this approach

# Building Automation System - BIM Integration Using a Linked Data Structure

Caroline Quinn[1], Ali Zargar Shabestari[3], Sara Gilani[1], Tony Misic[2], Marin Litoiu[3], J.J. McArthur[1]*

[1] Department of Architectural Science, Ryerson University

350 Victoria St., Toronto, ON, M5B 2K3

[1] Department of Computer Science, Ryerson University

350 Victoria St., Toronto, ON, M5B 2K3

[3] Department of Electrical & Computer Engineering, York University,

LAS 1012M, 4700 Keele St., Toronto, ON, M3J 1P3

*corresponding author: jjmcarthur@ryerson.ca

(phone : +1 416-979-5000 x554082 fax: +1-416-979-5353 )

## Abstract

Buildings Automation Systems (BAS) are ubiquitous in contemporary buildings, both monitoring building conditions and managing the building system control points. At present, these controls are prescriptive and pre-determined by the design team, rather than responsive to actual building performance. These are further limited by prescribed logic, possess only rudimentary visualizations, and lack broader system integration capabilities. Advances in machine learning, edge analytics, data management systems, and Facility Management-enabled Building Information Models (FM-BIMs) permit a novel approach: cloud-hosted building management. This paper presents an integration technique for mapping the data from a building Internet of Things (IoT) sensor network to an FM-BIM. The sensor data naming convention and time-series analysis strategies integrated into the data structure are discussed and presented, including the use of a 3D nested list to permit time-series data to be mapped to the FM-BIM and readily visualized. The developed approach is presented through a case study describing the scalability of the approach to integrate a building BAS system with a BIM. The resultant data structure and key visualizations are presented to demonstrate the value of this approach, which permits the end-user to select the desired timeframe for visualization and readily step through the spatio-temporal building performance data.

## 1    Introduction

Parametric design tools, Building Information Models (BIMs), digital fabrication, and virtual construction scheduling generate a wealth of digital data on the built environment through the design and construction phases. As a building is put into operation, the volume of this information increases exponentially and Facility Management-enabled BIMs (FM-BIMs) are of specific value in this digital context, having demonstrated time- and cost-savings benefits for a breadth of facility management activities (Arayici, Onyenobi and Egbu 2012, Bryde, Broquetas and Volm 2013, Volk, Stengel and Scultmann 2014, Love, et al. 2014). When Computer Aided Facility Management (CAFM) data is integrated into a BIM, it supports operational uses such as utility cost reductions, comfort management, space optimization, improved inventory management, and energy management (Love, et al. 2014). Over time, the volume of this information becomes extremely large and cumbersome, as the operational data requires a high level of effort for BIM maintenance and integration (Volk, Stengel and Scultmann 2014). While seen as highly promising in facility management (Bryde, Broquetas and Volm 2013, Volk, Stengel and Scultmann 2014) BIMs are still used as standalone information systems by building stakeholders. By integrating systems (ex: IoT sensor network and FM-BIM), and working towards aggregating data in a single model, negative effects of standalone, information systems can be minimized. A central FM-BIM facilitates the integration of all building data sources, to be used through the building project and operational lifecycle.

This paper presents a database architecture integrating IoT sensor data to an FM-BIM, with a specific focus on the pre-processing, effective storage, and mapping of time-series data. A case study of the proposed integration on a university building is presented to demonstrate how each of the sensor data streams can easily be mapped to fields within the FM-BIM using Dynamo, a Visual Programming Language (VPL). Sensor data values for desired time frames can be selected by the user in the VPL, permitting a variety of methods to navigate and visualize this data. The research offers a database architecture which could act as the data source for cloud based Smart and Continuous Commissioning (SCCx), an accessible portal to the BAS sensor data typically hidden within a proprietary system, and data analytics and visualization strategies. Together, these facilitate FM-BIM Digital Twins that could be used by Facility Managers (FMs) to test and plan maintenance projects, monitor, infer, and predict building conditions as a function of outdoor weather conditions and system settings. These insights can then be used within a SCCx approach to optimize building performance in near-real time.

## 1    Literature Review

Three challenges must be overcome to link an FM-BIM with live IoT data. First, a suitable architecture to support the operational IoT data – both static and dynamic - must be defined (Ramprasad, et al. 2018). Second, the integration method (Preidel, Daum and Borrmann 2017) must be determined. Finally, the IoT data to integrate must be identified and clearly defined (Pishdad-Bozorgia, et al. 2018).

## 1.1 Architecture for IoT – BIM Integration

A challenge in integrating IoT data to an FM-BIM is their heterogeneity. In general, FM-BIMs contain primarily semantic, geometric, and topographical (static) data, while IoT sensor data streams are time-series (dynamic) in nature (Ramprasad, et al. 2018, Corry, et al. 2015). The most common integration technique for static and dynamic data is referred to as *linked data* (Li, et al. 2018, Kaed, Leiday and Gray, Building Management Insights Driven by a Multi-System Semantic Representation Approach 2016, Hu, et al. 2018). This approach stores each building data source separately, effectively creating a data lake that is accessible through a common data management system (Pauwels, Corry and O'Donnell 2014). In the case of this research, static spatial data is hosted in the BIM and dynamics in the IoT datastore. Frequently this approach is used with links to decoupled ontology (static semantic data) and IoT time-series (dynamic) database sets; using a query processor to make this data accessible to applications (Hu, et al. 2018, Couloumb, et al. 2017). This approach will be referred to as *ontology linked*. In other instances, standardized naming formats can be used to create a link directly between the IoT and BIM data points (Damm 2013). This approach will be referred to as *directly linked*.

In an *ontology linked* approach an ontology acts as a proxy to link IoT and Building Automation System (BAS) data and BIM. This approach has previously been explored, this includes with the use of the Semantic Sensor Network (SSN) ontology (Kučera and Pitner 2016) and a custom Building Automation and Control Systems (BACS) ontology (Terkaj, Schneider and Pauwels 2017). The SSN ontology is a "generic language to describe sensor assets" (Neuhaus and Compton 2009) and has been used to describe BAS data with semantic tags such as "building", "room", and "device" and relate them using the "hasPhysicalProperty", and "sensingMethodUsed" attributes so that it can be linked to the BIM (Kučera and Pitner 2016). Some of the SSN tags are custom extensions, needed to create the link between the SSN ontology and BIM, additionally, a large amount of semantic data needs to be specified manually for each point in the BAS. Further, because SSN is a broad approach rather than one developed specifically for the building domain, there is no functionality to map SSN ontology data to the BIM as represented in open format (Industry Foundation Class, IFC). As such, an initial mapping schema must also be created to support this linked data approach. The BACS ontology is a new ontology developed by reusing existing ontologies such as SSN, Building Topology Ontology (BOT), Sensor Observation Sample and Actuator (SOSA), a fragment of the ifcOWL ontology (ifcmr), and others. Within BACS, BOT is used to describe spatial building data such as "floor", "space" and "element", SOSA is used to describe "Sensors" and other "FeatureofInterest" within the IoT network, and ifcmr is used to represent sensors readings as "Values" (Terkaj, Schneider and Pauwels 2017). The resulting ontology is robust and covers semantic data required to link BAS, IoT, and BIM data. The reuse of existing ontologies resolves issues of using the SSN ontology alone, the SSN ontology does not need to be extended, and the inclusion of ifcmr indicates that linking the BIM IFC would be possible. However, the need to manually map ontology tags to data points remains unresolved currently.

Ontologies that have not been directly used to link BAS/IoT data to BIM but could conceivably be used for the application include Brick (Balaji, et al. 2016) and the Haystack Tagging Ontology (HTO) (Charpenay, et al. 2015) . The Brick ontology is specifically designed to describe building HVAC system data. It has a hierarchical design, where tags of each subsequent layer of the ontology provide more detail, for example a "Fan" is part of an "Air Handling Unit (AHU)", which is part of "HVAC", *etc.* (Balaji, et al. 2016), which is similar in structure to a traditional BAS. The availability of some locational data in the Brick ontology further makes it a good candidate to link BAS and BIM data. HTO expands the Haystack ontology (Quan, Huynh and Karger 2003), which has an embedded, but incomplete, tagging approach. HTO uses Semantic Web technologies (RDF, OWL, SPARQL) to address gaps in Haystack from an IoT perspective; notably the requirement for computational power at the IoT device, the incompatibility of the API with standard web approaches such as REST and JSON, and the lack of a formal representation that can limit scalability (Charpenay, et al. 2015).

The use of an ontology acting as a link proxy between BAS/IoT and BIM presents the issue of data redundancy. The ontology tags facilitating the link are related to spatial or contextual building information, however FM-BIMs already store this data inherently. Ontology semantic data is therefore redundant in this use case. A *Directly linked* approach, linking *BAS*/IoT to BIM using standardized naming formats is much simpler. Within the construction sector, the Construction-Operations Building information exchange protocol (COBie) (East and Carrasquillo-Mangual 2013) was developed to establish such a standard naming format to facilitate the integration of Computerized Maintenance Management System data to BIM. Research aiming to leverage this format to automatically create these links and circumvent manual work had limited success due to interoperability issues with real world applications (Pishdad-Bozorgia, et al. 2018). Despite this limitation, there have been industry examples where systems conforming to a parsable and interpretable naming convention enable direct linking of data. For example, the taxonomy present in a data point tag can indicate the system, and equipment a data point is related to within the BIM, however spatial data must be represented in look up tables (Damm 2013). This *directly linked* data architecture requires no manual tagging; instead, the integration and connection between BAS/IoT data stores and other datasets define the resultant input data structure. The Open Messaging Interface (O-MI) and the Open Data Format (O-DF) have been used for a direct linking approach, and although there remains duplication of spatial data with this approach it would be well suited for direct linking if a standardized naming convention has not been used to describe data points (Dave, et al. 2018).

Table 1 demonstrates a comparison of *ontology and directly linked* approaches. It can be concluded that a *directly linked* approach is of value for BAS to BIM integration due to its ease of implementation when a standardized naming convention has been used for IoT data points. An *ontology linked approach* is both time consuming to implement and is restricted to semantic data concepts represented in the ontology. While a *directly linked* approach does not support semantic web technologies without the duplication of spatial data other aspects of the linked data ecostructure, such as spatial data serialized to IFC, can benefit from this approach.

Table 1: Integration Approach Comparison

| | Strength | | Weakness | | | |
|---|---|---|---|---|---|---|
| | Supports semantic web technologies | Previously used in IoT-BIM integration | IoT data must be named with standardized naming convention | Duplication of spatial semantic data | Ontology Extension and mapping schema required to support linked data | Manual implementation required to facilitate linked approach |
| Directly Linked Approach | | | | | | |
| FuseForward (Damm 2013) | ●* | ● | ● | ● | | |
| Otaniemi3D (Dave, et al. 2018) | ●* | ● | | ● | | ● |
| Ontology Linked Approach | | | | | | |
| SSN (Kučera and Pitner 2016) | ● | ● | | ● | ● | ● |
| BACS (Terkaj, Schneider and Pauwels 2017) | ● | ● | | ● | ● | ● |
| Brick (Balaji, et al. 2016) | ● | | | ● | ● | ● |
| Haystack (Charpenay, et al. 2015) | ● | | | ● | ● | ● |

*These technologies are not yetsupported in referenced publications

## 1.2 Integrating to FM-BIM

Central to both architectures approaches –*directly linked* and *ontology linked* – the use of a query processor is required to retrieve the appropriate subset of summarized time-series data to be imported to the FM-BIM. This is because a link exists between IoT and BIM data points in either architecture, and a query processor is required to present summarized IoT rather than raw time-series data, the value of which is highlighted in this research. Two query processors have been developed to address this integration using an *ontology linked* approach: HodDB (Fierro and Culler 2018) and ForTÉ (Kaed and Boujonnier, FOrT ´E: A Federated Ontology and Timeseries query Engine 2017). HodDB contains a query processor, which accepts SPARQL queries and processes by first accessing Brick ontology data is stored in a LevelDB, and then the time-series data stoered in a BTrDB (Fierro and Culler 2018). Query time is very fast using HodDB due to the use of a specific database and ontology. ForTÉ, is a more flexible query processor, while it also uses SPARQL, it can be used for any database types and ontology (Kaed and Boujonnier, FOrT ´E: A Federated Ontology and Timeseries query Engine 2017). Selecting the appropriate query processor when integrating IoT time-series data to an FM-BIM if using and *ontology linked* approach has large implications on the overall integration time. This is especially true when a user is requesting multiple visualizations within the FM-BIM during a single session, as can be expected in the use of an FM-BIM (Fierro and Culler 2018).

As noted previously, building information from IoT output can be integrated to FM-BIMs using the *directly linked* data architecture, where queries are written using the known naming convention to extract summary time-series data from the IoT database. Preidel et al. (2017)

explored the use of VPLs for the purpose of querying a *directly linked* data architecture and developed two task-specific VPL query engines: (1) QL4BIM for querying BIM semantic data represented in IFC; and (2) VCCL for code compliance checking. Neither of these are optimized for querying IoT summarized data; however, VPLs are easily learned and implemented by users and are well-integrated into some BIM software, for example Dynamo for Revit.

Once summary time-series data is available, VPLs such as Dynamo for Revit can be used to integrate time-series IoT and CAFM data to BIM, for example (Khaja, Seo and McArthur 2016, Bortoluzzi, et al. 2019). Central to these approaches are the parsing and cross-referencing of unique identifiers for individual elements that can be matched between datasets. This is critical when using a *directly linked* data architecture; in such approaches, an integration method based on the naming convention is uniquely responsible for mapping uniquely identified data points. Conversely, in the *ontology linked* approach the semantic tags associated with the time-series data also aid in mapping IoT data points to the BIM. An alternative approach to using VPLs in the *directly linked* approach is an Extract Transform Load (ETL) tool, which has been developed to handle multiple data sources to populate a target data user such as an FM-BIM. An FM-BIM specific ETL has been developed for this purpose however it did not consider time-series FM data sources (Kang and Hong 2015). Both approaches also require the use of communications protocols to facilitate the transfer of IoT data to cloud-based storage. Previous *directly linked* approaches have used a streaming device connected to the internet via ethernet connection, in this approach IoT devices are connected to the streaming device over a Virtual Private Network (VPN) from collecting IoT data. The benefit of this approach is the security, speed, and reliability of data transfer; however, if this hardware connection does not exist wireless approaches are required. Dedicated 5G networks have not been tested due to the unavailability of network hardware for testing. Dedicated 4G networks *can* be used when ethernet connections are not available.

## 1.3    Summary Data for Integration

The need for appropriate "data resolutions" refers to the need for summarized time-series data with larger time granularity within an FM-BIM as opposed to the raw, fine granularity collected by sensors, actuators, and meters (Gerrish 2017). Independent of the approach used – whether *ontology linked* or *directly linked* – summarized data is required based on the desired visualizations and stored in the time-series database along with the raw IoT data. For cases where a *directly linked* approach is used, the summary data point tag would also need to be available to the query engine, whereas if using an *ontology linked* approach, the summary point nodes would have to be available in the ontology to permit summary data discovery during time-series data queries. An example of the former is the work of Gerrish (2017), who proposed a *directly linked* data architecture approach to integrating IoT data to BIM using a built-in IFC node "PerformanceMetric", noting that its connection to IFC sensor nodes would facilitate direct import to the BIM, though this is not yet possible in the IFC schema. An example of the latter is the Brick ontology, which was designed to permit the inclusion of summarized data tags by creating these semantic concepts within the ontology (Balaji, et al. 2016).

Looking beyond sensor data streaming, there has been a significant body of work undertaken to identify information required by FM to support building operations (Bryde, Broquetas and Volm 2013, Volk, Stengel and Scultmann 2014, Kassem, et al. 2015). In one example (Kiviniemi and Codinhoto 2014), IoT data from the Building Management System (BAS) was used to support building operations and maintenance. The integration and visualization of complex time-series data, for example time-series visualization within 4D BIM, has been shown to allow FM users to synthesize more complex data than was previously possible (Motamedi, Hammad and Asen 2014). Motamedi et al. (2014) demonstrated the value of visualizations using icons and symbols, the inclusion of 3D building components, and color-coding zones/spaces to support root cause failure detection. A general integration of IoT data to BIM by Dave et al. (2018) provides visualizations in the form of data point trend lines, 2D models with heat maps, 3D models with data point location, as well as geographical map views. While some of these visualizations can be viewed in different time granularities, they have been created for the use of end users and researchers, as opposed to specifically for Facilities Management.

## 1.4    Research Gap

The use of standardized naming conventions to support the directly linked data approach offers significant benefits for IoT-BIM integration, as evidenced by the literature review, reducing data redundancy and facilitating implementation by avoiding the ontology mapping step. The use of consistent data descriptions between the building and BIM further support direct transfer, reducing the work required for the implementation of IoT to FM-BIM. Previous research (Ramprasad, et al. 2018, Khaja, Seo and McArthur 2016, Bortoluzzi, et al. 2019), demonstrates the value of Dynamo to integrate time-series data stores using a standardized naming convention when using a directly linked data approach. However, this approach suffers from a lack of optimization for querying and the visualization possible when using an ontology linked approach. Further, time-series data summarization to the appropriate granularity and navigation over a time window are required to create visualizations supporting higher-level analytical decision making for FM. An alternative approach to circumvent the former issue while providing this navigation through time is presented herein.

Despite the significant research to date on the topic of IoT-BIM integration, there remains a paucity of literature regarding data streaming from sensor networks, particularly with regard to the mapping of time-series summary information to an FM-BIM (Kassem, et al. 2015). This research paper aims to fill this gap by presenting a complete IoT data acquisition, management, and BIM mapping integration method in the form of a *directly linked* data architecture as executed for two case studies. In the first, a lab-scale Arduino and Pi-based sensor network collects data and streams it to a cloud environment. The second case study used a modified streaming approach, but the latter streams approximately 10,000 Building Automation System (BAS) data points from a large (25,000m$^2$) building and pre-processes it to support future analytics. Both case studies use the linked data approach for BIM integration, with streaming and pre-processing adjusted for the larger case study to adapt tospecific site conditions.

## 2 Approach

An IoT sensor data integration to FM-BIM requires three data processing components: (1) a data acquisition and ingestion system; (2) batch analytics, and (3) an integration engine. The data acquisition and ingestion system collect the raw sensor data and transfers it to a cloud database. The batch analytics summarize the data using a set of defined business rules to permit faster data querying, and the queried data is then mapped to the FM-BIM using the integration engine. Figure 1 shows how these processes act on the various system elements (sensor points, database of historical records, and the FM-BIM). This approach follows from the findings cited in the literature review.

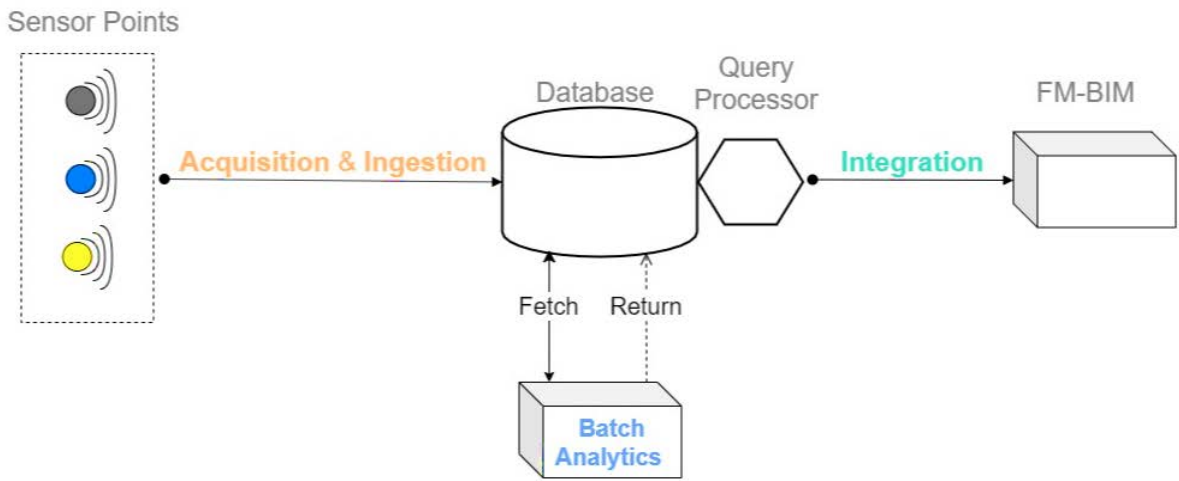

*Figure 1:* System architecture for IoT data streaming to FM-BIM

In order to provide an efficient and flexible architecture, minimal processing occurs during acquisition and ingestion, while cloud-based environments are used for data pre-processing and processing given their computational power. Data point values are given tags, unique identifiers stored as string of characters in the database system, organized in a hierarchical schema where the building is broken into successively more granular components. The hierarchy for a building is shown in Figure 2. This relates to the point naming convention: *BuildingID.SysID.BASID.PointID*. The system context is defined as follows. For monitored equipment, the SysID is the building system containing the equipment, for example, the chilled water system, and BASID is the actual equipment being controlled, for example, a chiller or chilled water pump. For room-based points, SysID is assigned "RM" and BASID is the room number. Each sensor point type is then assigned a PointID. The resultant naming schema concatenates the building, system, element, and point names to create both a machine-parsable and human interpretable ID, allowing FM users to interpret the relevance of each IoT point to the building.

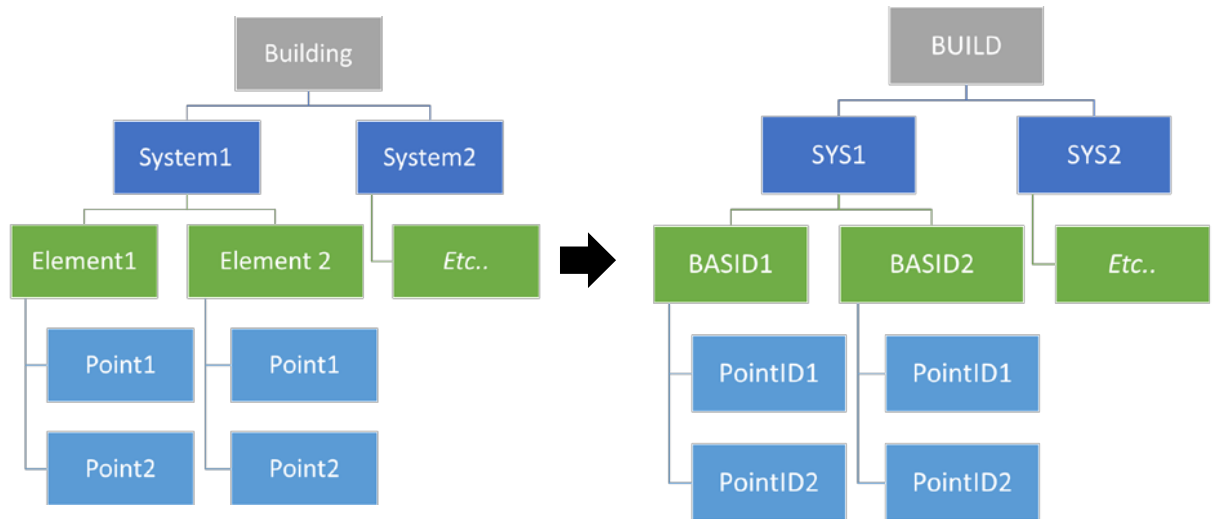

*Figure 2: Database Hierarchical Schema – general (left) and controls-specific implementation in this paper (right)\*

## 2.1 Case Study #1: Local Sensor Network

A living lab test cell has been established within a single faculty office, which incorporates a local sensor network. Sensors measuring occupancy, lighting state, door position, and HVAC are integrated with an Arduino Mega 2560 and streamed via an Ethernet connection to a private cloud. Other systems in this office but not discussed in this paper measure ambient temperatures in the office and adjacent spaces (direct cloud streaming) and thermocouples for surface temperature measurement. Figure 3 shows setup of the IoT sensor network in the office living lab and a sample of .csv data to be mapped to BIM.

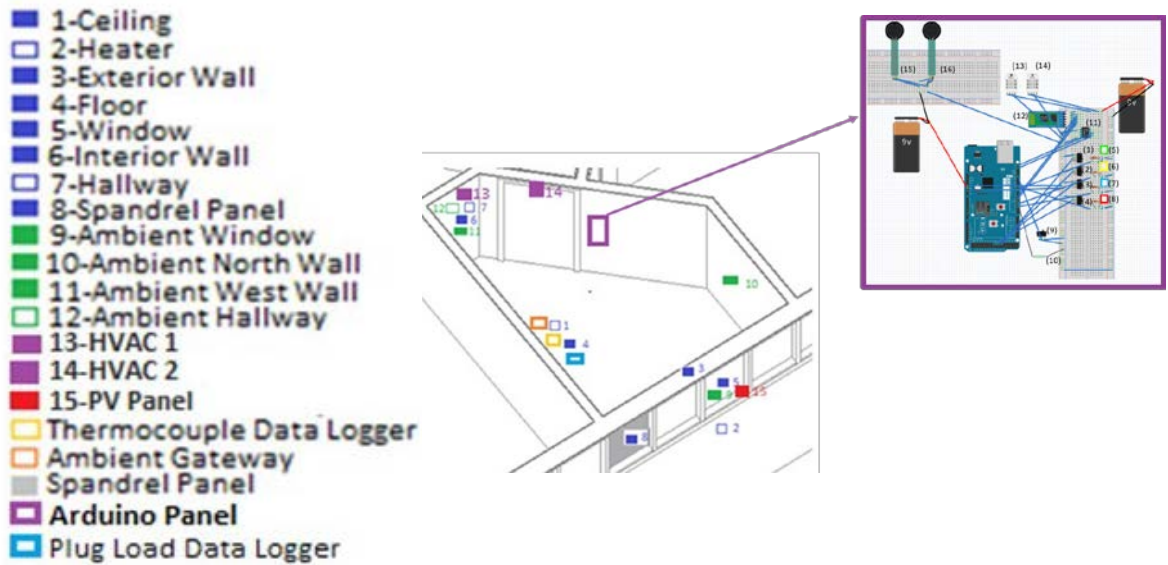

| | A | B | C |
|---|---|---|---|
| 1 | timestamp | ARC-0309 | ARC-0314 |
| 2 | '14-Mar-2019 19:44:00' | 0.2,0.2,0.1,12.1,36.2,20.8,34.6,24 | 0,0,0,10.4,33.1,20.4,33.8,22.6 |
| 3 | '15-Mar-2019 19:44:00' | 0,0,0,18.7,41.9,18.9,35.3,23 | 0,0,0,10.6,56.5,18.9,43.7,23.5 |
| 4 | '16-Mar-2019 19:44:00' | 0,0,0,12.4,26.3,20.9,27.7,23.3 | 0,0,0,10.9,36.5,20.2,34.6,23.5 |
| 5 | '17-Mar-2019 19:44:00' | 0,0,0,11,21,21.2,23.8,23.6 | 0,0,0,10.9,45.4,19.7,38.4,23.5 |
| 6 | '18-Mar-2019 19:44:00' | 0.1,0.1,0.1,10.9,26.6,20.6,27.4,23.6 | 0,0,0,10.9,48.9,19.1,40,23.7 |
| 7 | '19-Mar-2019 19:44:00' | 0.3,0.3,0.2,13.4,30.4,20,29.7,23.4 | 0,0,0,11,21,21.2,23.8,23.6 |

**Legend:**
- 1-Ceiling
- 2-Heater
- 3-Exterior Wall
- 4-Floor
- 5-Window
- 6-Interior Wall
- 7-Hallway
- 8-Spandrel Panel
- 9-Ambient Window
- 10-Ambient North Wall
- 11-Ambient West Wall
- 12-Ambient Hallway
- 13-HVAC 1
- 14-HVAC 2
- 15-PV Panel
- Thermocouple Data Logger
- Ambient Gateway
- Spandrel Panel
- **Arduino Panel**
- Plug Load Data Logger

*Figure 3: Sample of IoT data streaming (top) office living lab set up (bottom) and pre-processed daily average CSV rows*

## 2.2 Case Study #2: Large Mixed-Use Building

A Health Science (DCC) complex on campus at Ryerson University in Toronto Ontario with 8 academic floors,18 residence floors, a podium, and 4 levels of underground parking has been designed as a Living Lab (Figure 4). Approximately 10,000 data points are streamed from this facility in near-real time, triggered by changes of value above a defined threshold, to an ElasticSearch Cluster.

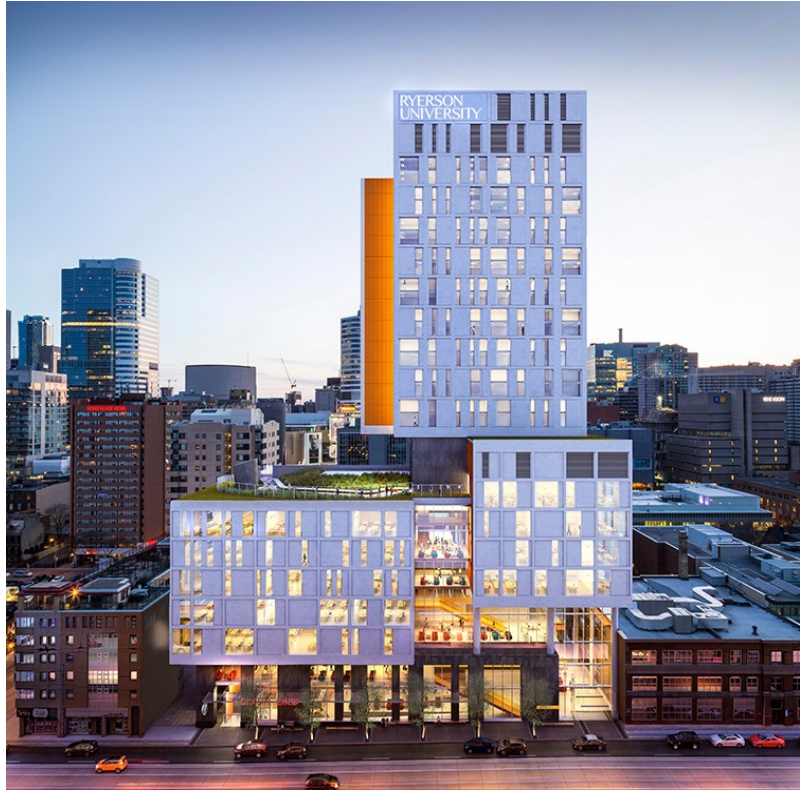

*Figure 4: Daphne Cockwell Complex at Ryerson University* (Perkins + Will 2019)

## 3    Data Acquisition and Ingestion

### 3.1    Case Study #1 Implementation

*Controllers* accept sensor data and periodically push it to a local buffer that synchronizes the time-series data. This is a more efficient approach than synchronizing each data point stream directly in the *buffer*, accommodating different measurement timescales while avoiding the need for multiple network connections. A *streamer* then pulls a larger set of synchronized data from the buffer and inserts in the cloud datastore representing a longer timeframe. The controller will allow data to be streamed more quickly from IoT data point to the FM-BIM, giving FM users a near live view of building operations (Ramprasad, et al. 2018).

Recommendations for software and hardware to be used at each step of the IoT data acquisition and ingestion as well as a schematic of the process can be seen in Figure 5.

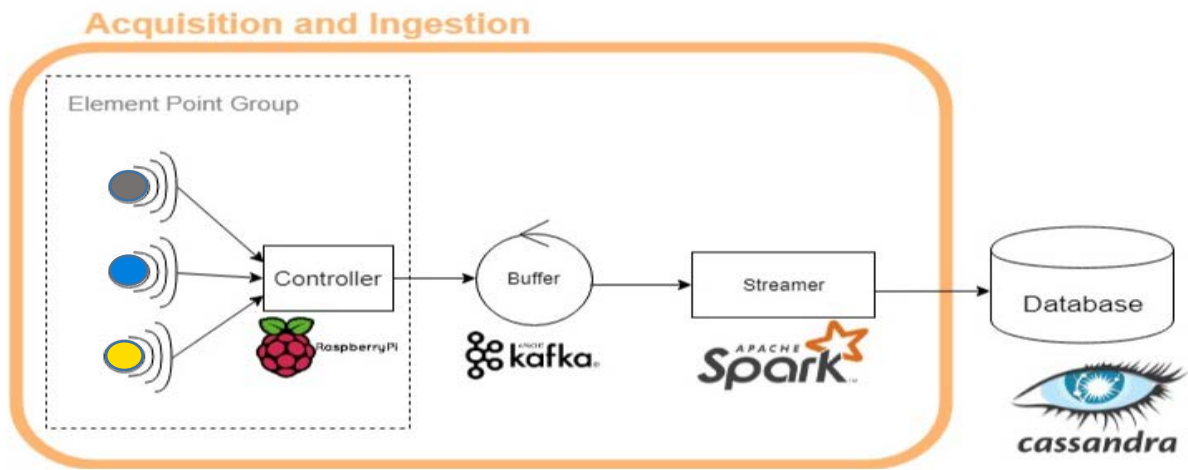

Sensors with unique Identifiers: **BUILD.SYS.BASID.POINTID**

*Figure 5: Data acquisition and ingestion process*

Because a hierarchical schema is used for tagging data points in both the time-series data stores and FM-BIM, a variety of timestep granularities can be supported for visualization. This is achieved by aggregating run time queries using SPARK, a process further described in section 3.2. Because the time-series database point data is partitioned by element, point, and month, individual points (POINTIDs) can be decoupled from their element groups (BASIDs), *i.e.* a data row can be created for individual *BUILD.SYS.BASID.POINTID* series to permit batch analytics to be run in parallel, increasing computational efficiency. It should be noted that in this approach row size restrictions of a NonSQL database such as Cassandra forced the point data be partitioned by month. Figure 6 visualizes IoT data storage within a Cassandra a nonSQL database. Time-date stamps are not used in the Cassandra database; instead, a count value is used to number each timestep from the beginning to the end of the month. Because time is represented on this scale, it increases the importance of the buffer to synchronize time-series data before pushing to the streamer.

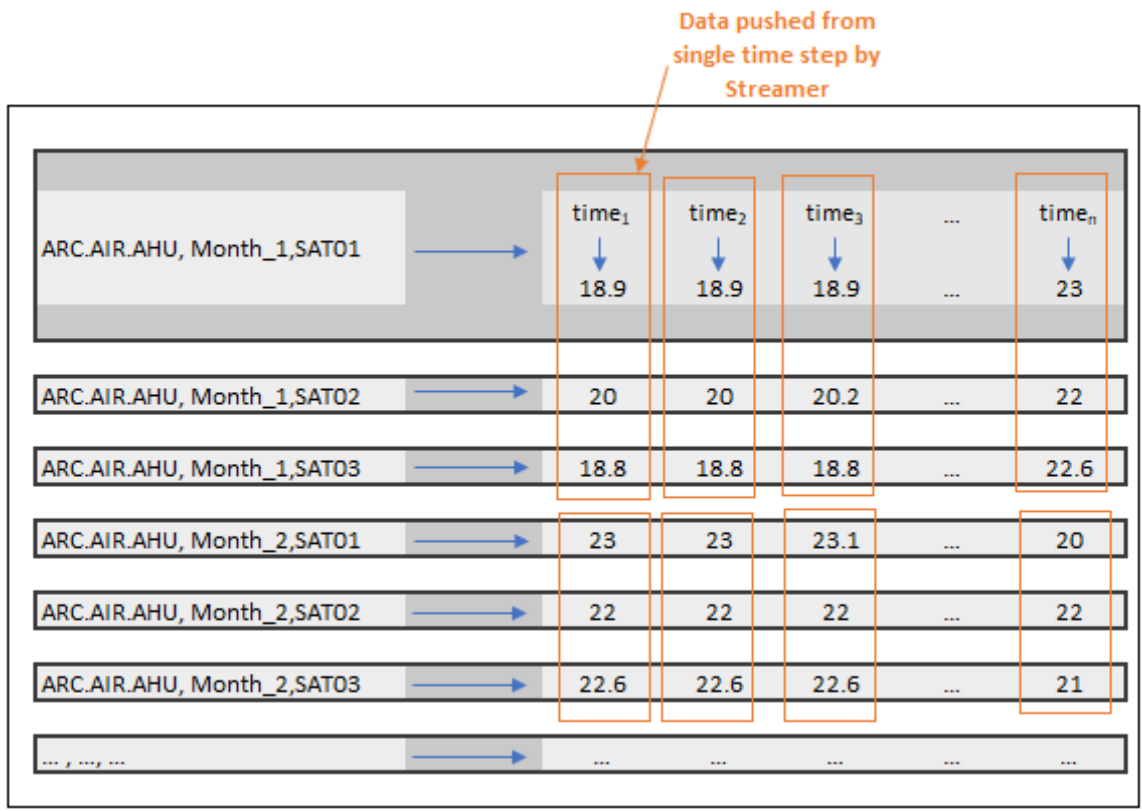

*Figure 6: IoT data storage in Cassandra*

## 3.2   Case Study #2 Implementation

There are two components in the data acquisition system: the BAS, and the software used to extract the BAS data and export it in a non-proprietary format. BAS systems vary by vendor, but typically follow the standard architecture described in Section 3.

BAS data is output to a physical machine located on the BAS ethernet network.Because overloading the BAS network with queries from the streaming system must be avoided as this could result in a lag in controls or at worst a network failure, a "push" output is required from the system when a change of value (COV) above a defined tolerance is recorded for each point. These new points are then "sniffed" from the network and recorded in the desired format. Unless this is performed downstream of each field controller, i.e. where BACNet IP is used for communication with individual equipment, this will be encoded in the proprietary BAS system and require coordination with the controls vendor to output these values in parsable text format. Because of the number of NAE devices in the building, the researchers collaborated with the controls vendor (Johnson Controls Inc.) and their subconsultant (AFDTek) to develop a 'back door' in the proprietary system that provided this output. At each COV event, the timestamp, full-context pointID (which is named by this system based on the point network location in the format

*NAEDeviceID.TrunkID.FieldControllerID.PointType* and the new value was embedded into the *Data* field (TCP body message) as text in a TCP packet (Figure 7). The script is then assigned a predetermined IP address and port and that location information is input to the BACnet UI. The remainder of packet information is populated using standard TCP protocol rules (Dordal 2014).

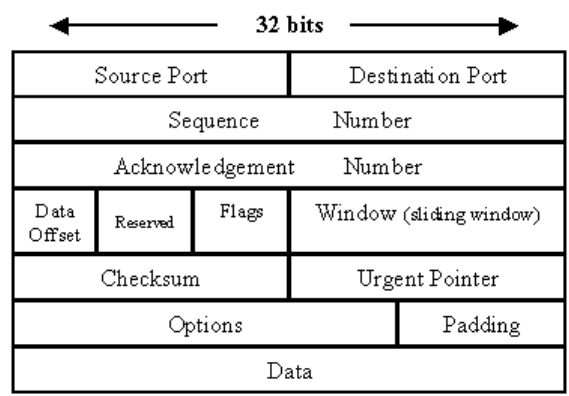

*Figure 7: TCP packet structure* (Salomon 2006)

Completed TCP packets are streamed to their respective packed-coded locations via the Python server to an ElasticSearch (ES) cluster. This is shown schematically in Figure 8.

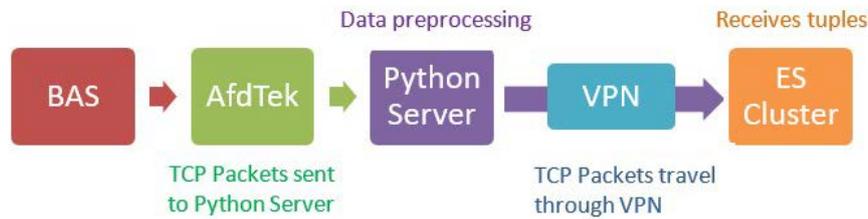

*Figure 8: Schematic of data streaming architecture from BAS to ES Cluster.*

The COV data sent by the BAS network devices to the Python server script, extracted from the TCP body message, and parsed using the *.split* and *.strip* functions in Python to populate the {timestamp, device ID, point ID, value} tuple. Once the tuple has been populated, the script attempts to write the data into the ES cluster index. Three means are used to avoid data loss. First, the IDs of the documents in the index are an Md5 hash (Salomon 2006), a type of cryptographic key, is generated at each push attempt, minimizing the chance of duplicate ID's and avoiding non-duplicate data writing. Second, a backup file is written to each time a push attempt fails. In the process of writing from the file, indexing to ES is verified. This technique eliminates the loss of data during a downtime of the ES cluster or a client/server error. Finally, the entire server/client system is run over TCP/IP, which will guarantee information security and will not receive any incomplete data.

Cybersecurity is enforced with a secure VPN, which connects the physical machine to the ES cluster. This VPN connection is configured such that only the confirmation of TCP packet arrival may be sent back to the Python server to minimize the risk of cyber-intrusion. A sample of data communication to and from the ES cluster is presented in Figure 9.

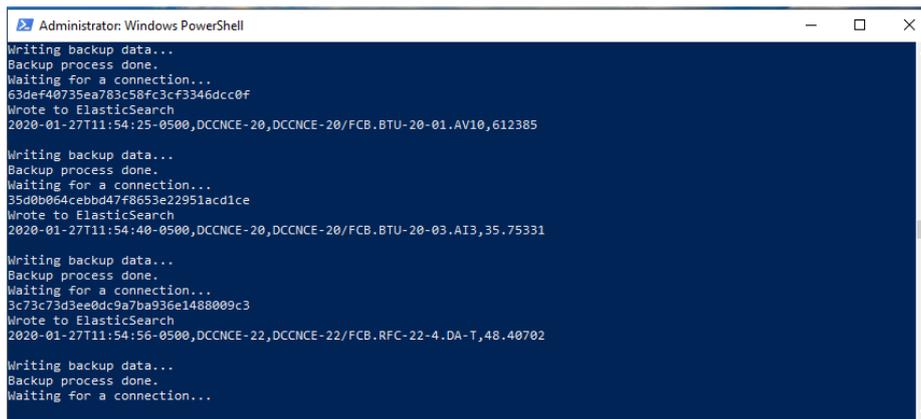

*Figure 9: Window showing data streaming to ES server via VPN*

It is acknowledged that a direct connection of the BAS head-end to the cloud would preferable from a reliability standpoint as this would eliminate several potential points of failure – for example risk of disconnected cables, power failure at this intermediate computer, or system restarts – as well as lag associated with the TCP confirmation process. However, the facility management and cybersecurity officers at the university required additional security in order to obtain permission to connect to the BAS network.

### 3.2.1 Pre-processing

Figure 10 shows a sample of the raw data as received by the ES Cluster, prior to the structuring and pre-processing, which is required to enable efficient storage and use by future analytics algorithms.

| Time | asset_name | meter_name | reading_value | _id |
|---|---|---|---|---|
| January 29th 2020, 00:03:16.000 | DCCNAE-03 | DCCNAE-03/FC-1.CAV-3-10.T | 21.93143 | d2d45c4f1d5355ec20b0b36936c60546 |
| January 29th 2020, 00:02:12.000 | DCCNCE-28 | DCCNCE-28/FCB.HTH-28.AI-1 | 1005.604 | 03b0095e7d75efec02d0e3046cd3e561 |
| January 29th 2020, 00:01:57.000 | DCCNAE-04 | DCCNAE-04/FC-1.CAV-4-36.Q | 441 | e96043a966c8e30f4602902004b9edd |
| January 29th 2020, 00:01:40.000 | DCCNAE-06 | DCCNAE-06/FC-1.CAV-6-42.Q | 514 | 04fb4c5fddee431e37a4dff4addd02933ec |
| January 29th 2020, 00:01:24.000 | DCCNAE-03 | DCCNAE-03/FC-1.CAV-3-17.HC-O | 70.0466 | 60bcc057e5169bcd4a85e4cadc9de58c |
| January 29th 2020, 00:00:07.000 | DCCNAE-05 | DCCNAE-05/FC-1.CAV-5-44.Q | 206 | 3be40b419c69a49582d70101876dbea6 |
| January 29th 2020, 07:59:45.000 | DCCNAE-04 | DCCNAE-04/FC-1.CAV-4-27.HC-O | 46.05806 | e921a555ace975573705d335bf6b50a6 |
| January 29th 2020, 07:59:30.000 | DCCNAE-06 | DCCNAE-06/FC-1.FCU-6-2.Q | 623 | 562ae031cf0b00632b049b8751a106f |
| January 29th 2020, 07:59:15.000 | DCCNAE-03 | DCCNAE-03/FC-1.CAV-3-17.Q | 629 | b163b15f20a2970464e2619cbd0931ea |
| January 29th 2020, 07:58:04.000 | DCCNAE-05 | DCCNAE-05/FC-1.CAV-5-05.Q | 246 | ff60b7035ee2b034770a97111c674d0c |
| January 29th 2020, 07:57:48.000 | DCCNAE-04 | DCCNAE-04/FC-1.CAV-4-27.HC-O | 46.51545 | e045da04318bddade84ffba476b35a32 |
| January 29th 2020, 07:57:33.000 | DCCNAE-07 | DCCNAE-07/FC-1.CAV-7-31.T | 22.04713 | d39e6007bfcc3e1e60b6b2cae47ea7ad |
| January 29th 2020, 07:57:16.000 | DCCNAE-03 | DCCNAE-03/FC-1.CAV-3-39.Q | 671 | 72fcfea908bd0f07fa319fe41c7ea092 |
| January 29th 2020, 07:57:00.000 | DCCNAE-03 | DCCNAE-03/FC-1.CAV-3-14.Q | 437 | e5d01451dd5c4bc16dfe56b7d52337b9 |
| January 29th 2020, 07:55:46.000 | DCCNAE-05 | DCCNAE-05/FC-1.CAV-5-44.Q | 202 | 4d49255eef3c6f6e683d61d9b0e0cade |
| January 29th 2020, 07:55:29.000 | DCCNAE-01 | DCCNAE-01/CARPA_L1 DACnet IP1.CARPA_METER - EMP6.Analog_Values.AV-115 | 118.55 | e0c2e0d82c146ccbd9c7b0005f24e09e |
| January 29th 2020, 07:55:11.000 | DCCNAE-06 | DCCNAE-06/FC-1.CAV-6-42.Q | 510 | b5fbbf188b9b57b53b108a7f18c4078c |
| January 29th 2020, 07:54:54.000 | DCCNAE-03 | DCCNAE-03/FC-1.CAV-3-14.Q | 442 | 1c7553b0e77e6d493ded8a3564442265 |
| January 29th 2020, 07:53:30.000 | DCCNAE-05 | DCCNAE-05/FC-1.CAV-5-0A.SAF-SP | 20.37364 | 0daa3b0fe32a1901a7d64ed1516c0035 |

Figure 10: Raw storage of events as recorded on ES Cluster

The fundamental issue with this data is that while it provides the full network context for each point in a standardized and parsable format, which is appropriate for a *directly linked* approach, the equipment/zone and system context is missing despite being required to integrate the BAS IoT data to a BIM. Fortunately, the DCC complex BAS uses the point naming convention described in Section 3, which was cross-referenced with network locations using a lookup table (Table 2), thus permitting a consistency in nomenclature that facilitates parsing into the SQL database. Taking in the formatted data a pre-processing Python script maps each column from the given data to the system context point IDs, which map each sensor, actuator, and controller to the appropriate equipment and system as well as network location. These full data point names give this contextual information in a parsable format that can be mapped to a unique field in the BIM.

*Table 2: Lookup table for point context (selected rows)*

| Network Context PointID | System Context PointID |
| --- | --- |
| DCCNAE-01/FC-1.CAV-1-10.CLGUNOCC-SP | DCC.RM.DCC01-13.CLGUNOCC-SP |
| DCCNAE-01/FC-1.CAV-1-10.EFF-OCC | DCC.RM.DCC01-13.EFF-OCC |
| DCCNAE-01/FC-1.CAV-1-10.EFFCLG-SP | DCC.RM.DCC01-13.EFFCLG-SP |
| … | … |

The complex uses a standardized naming convention for BAS data points and is therefore appropriate for a *directly linked* approach to integrate BAS IoT data to a BIM. Data point names can be parsed for contextual building information that can be cross referenced to the BIM. The full data point names give all contextual information in a parsable format that can be mapped to a unique field in the BIM.

## 4    Batch Analytics

Once raw data has been acquired and streamed to the cloud-based database, time-series analytics can run in batch jobs to summarize data to single human readable metrics, which is desirable to simplify interpretation by FM users. The cloud-hosted environment is the appropriate place to run these analytics due to its stability, high capacity, and computational power, all of which are required to achieve the high complexity summary calculations. Figure 11 shows the general process of using Spark for batch analytics. In this process, the batch analytics are run on a Spark processor and apply business logic. The returned results are then returned to the database for insertion to summary tables on the database.

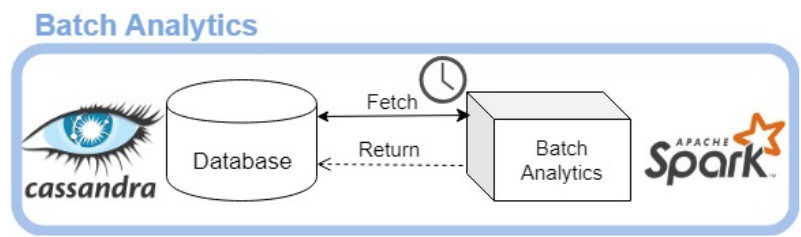

*Figure 11: Batch analytics process*

Appropriate time-series analysis methods depend greatly on the desired FM-BIM visualizations. Summarization functions, including but not limited to counts, averages, and min/max values can be integrated to an FM-BIM. In addition to selecting the most appropriate summarization function, the appropriate time granularity for reporting must also be selected. Time granularity refers to the amount of time be considered in a summary value, ex: one hours' worth of data, four hours' worth, one days' worth, one months' worth, etc. Figure 12 gives an example of temperature data being summarized from the sensor data – read in five-second intervals – to daily maximum values.

Time series

Summarized

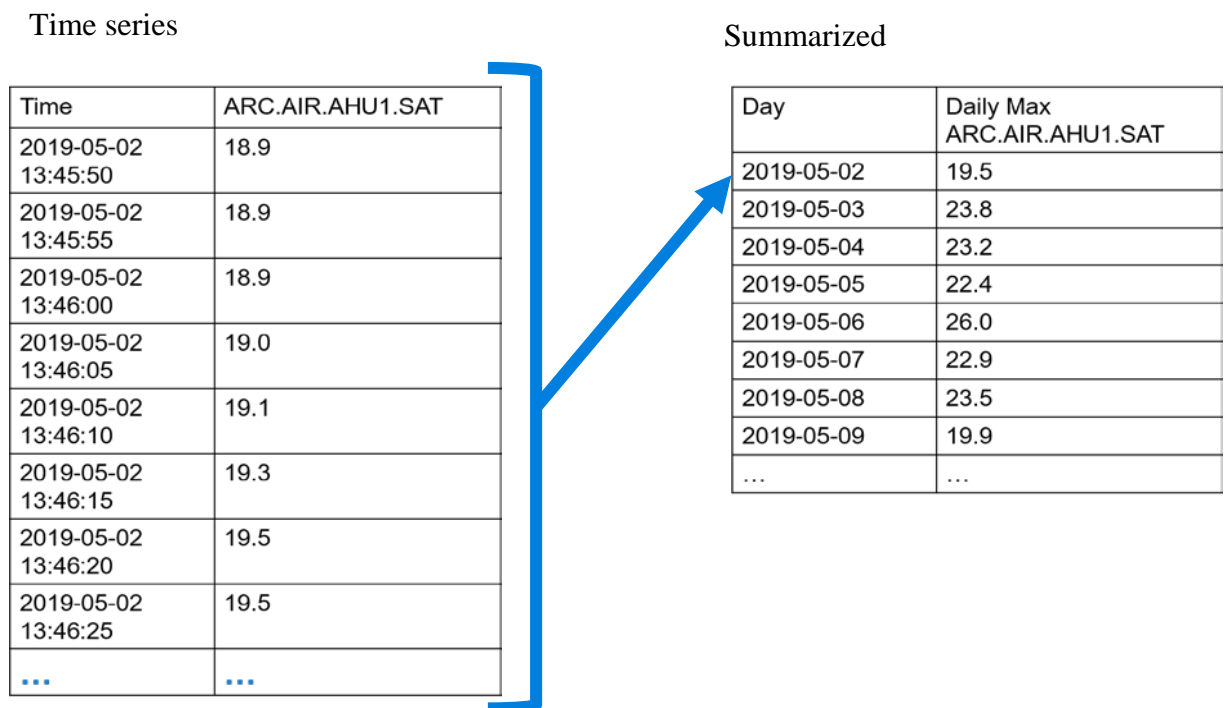

*Figure 12: Sample summary of temperature data*

Average values, desired for generalized result reporting rather than event counts, can be applied to any point data that are normally distributed. If sensor data is not normally distributed an average will not provide an accurate representation of conditions. Batch analytics will calculate the average value for each sensor at each hour, as well as identify the minimum and maximum value for the sensor within the hour.

Counts can be used as an effective means to visualize building deficiencies and therefore deficiency percent changes over time. Counts are an appropriate method of time-series analysis when point threshold values and time delays before a point value can be considered deficient are known. If counts are employed without time delays or specific threshold values are not known, they will transform data rather than effectively summarizing it. In the context of IoT sensor data, hierarchical alarms defined by a BAS system offer time delay and threshold values to be used as count conditions for system and building deficiencies. ASHRAE Guideline 36 mandates hierarchical alarms, where high-level alarms indicate only minor deficiencies, and low-level alarms indicate major system deficiencies. This approach can also be applied to time-series IoT data analysis so that building stakeholders can visualize system and room deficiencies in the FM-BIM.

The translation of raw timeseries data to a file ready for import to the BIM requires summarisations to be made. Summarizations are made for each BASID of the desired types. for example all FCU's. During this process, raw IoT data is summarized to values that can be visualised within a BIM. Table 3 shows a sample of BAS data captured for BASID's FCU-1-11, FCU-1-13, and FCU-1-15, this is how the data would appear after acquisition and ingestions once it is stored in a cloud-based database. Each of these BASID's represents a Fan Coil Unit (FCU) located withing the building. The POINTID then indicates specific sensors associated to the BASID, DA-T for example, refers to Discharge air temperature, CLG-O to cooling output, and Q to CO2 ppm. Batch analytics providing summarizations of values can include the minimum air flow value, number of times zone temperature was below or above zone setpoint, average zone temperature, and the number of times effective occupancy- as recognized by the FCU- was outside occupancy schedule and therefore the zone required unplanned conditioning. Tables 4-7 demonstrate some hourly summarization of raw data Column headings reference the POINTID for which the summarized value will be mapped to in the BIM.

*Table 1 : Sample raw BAS timeseries data*

| Timestamp | Data Point name | Value |
|---|---|---|
| **2020-01-14 22:09:54** | DCCNAE-01, DCCNAE-01/FC-1.FCU-1-11.DA-T | 19.41395 |
| **2020-01-14 22:09:54** | DCCNAE-01, DCCNAE-01/FC-1.FCU-1-13.Q | 484 |
| **2020-01-14 22:09:54** | DCCNAE-01, DCCNAE-01/FC-1.FCU-1-15.CLG-O | 27.90866 |
| **2020-01-14 22:09:54** | DCCNAE-01, DCCNAE-01/FC-1.FCU-1-15.DA-T | 17.71609 |
| **2020-01-14 22:09:54** | DCCNAE-01, DCCNAE-01/FC-1.FCU-1-15.Q | 417 |

*Table 2: Representation of the minimum supply airflow for FCU 1-15 over sequential time periods as processed for BIM mapping*

| Timestamp | DCC.RM.DCCO1-04/FCU-1.FCU-1-15.FULLSCALE-SAF |
|---|---|
| **2020-01-14 20:09:54** | 30.3 |
| **2020-01-14 21:09:40** | 31.0 |
| **2020-01-14 22:09:54** | 31.1 |
| **2020-01-14 23:09:50** | 31.1 |

*Table 3: Representation of the count of temperatures not meeting setpoint as processed for BIM mapping*

| Timestamp | DCC.RM.DCCO1-04/FCU-1.FCU-1-15.T-SP |
|---|---|
| 2020-01-14 20:09:54 | 4 |
| 2020-01-14 21:09:40 | 4 |
| 2020-01-14 22:09:54 | 3 |
| 2020-01-14 23:09:50 | 2 |

*Table 4 : Average zone temperature as processed for BIM mapping*

| Timestamp | DCC.RM.DCCO1-04/FCU-1.FCU-1-15.T |
|---|---|
| **2020-01-14 20:09:54** | 20.1 |
| **2020-01-14 21:09:40** | 19.0 |
| **2020-01-14 22:09:54** | 16.0 |
| **2020-01-14 23:09:50** | 16.6 |

*Table 7: Number of times effective occupancy as recognized by the FCU was outside occupancy schedule*

| Timestamp | DCC.RM.DCCO1-04/FCU-1.FCU-1-15.EFF-OCC |
|---|---|
| **2020-01-14 20:09:54** | 1 |
| **2020-01-14 21:09:40** | 0 |
| **2020-01-14 22:09:54** | 0 |
| **2020-01-14 23:09:50** | 0 |

## 5    Integration

The integration of the IoT data to the FM-BIM is applicable to both case studies and consists of three key steps, summarized in Figure 13.

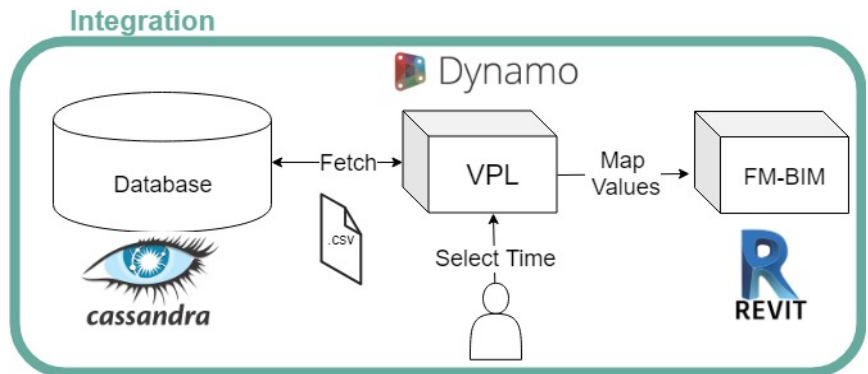

*Figure 13: IoT Sensor Database to FM-BIM integration process*

Within the VPL, there are three intermediate steps required to support FM-BIM mapping. First, a CSV file (format in Table 8) is passed to the VPL using a script. This file must be maintained from the summary tables, a separate file is required for each desired visualization summary type. This file is updated regularly as batch analytics are run with new incoming sensor data and saved with a fixed filename. This CSV file is a 2D list, columns headed with BASID and rows with timestamp inferred on querying the Cassandra datastore and reading the time value. Cells contain ordered comma delimited point values for the intersecting ID and timestamp. Second, the Dynamo script imports The CSV file, sorts it to match the FM-BIM element ordering, and creates a 3D list by converting the ordered point values for each BASID from a string to a list. Each index of the list contains a Revit element ID ordered list of point values. The user selects the desired time for display in the FM-BIM using a slider, which inputs to a function that filters the transposed 3D list using a python script and outputs a 2D further transposed list of sensor data for each BASID at the specified timestep. Finally, this list is mapped to the FM-BIM using the Dynamo *Element.SetParameterByName* node, updating the parameter in Revit of the element.

*Table 8: Sample CSV format for BAS mapping to BIM (truncated)*

|                       | BASID1        | BASID2        |
| --------------------- | ------------- | ------------- |
| **2019-04-24 17:00:00** | 0.23, 22.3, 5 | 0.20, 24.3, 1 |
| **2019-04-24 16:00:00** | 0.22, 22.3, 0 | 0.23, 24.2, 1 |

These CSV files can be used to map times series data such as averages or counts. A time slider can be used to select data for mapping and navigate over a desired time frame, for example, a 72-hour window with hourly average values, or a monthly window with daily averages values, for a given type of data point. These sliders control the FM-BIM visualization, providing FM users a visual and interactive platform to interpret sensor data. Figure 14 shows the Dynamo implementation for hourly navigation.

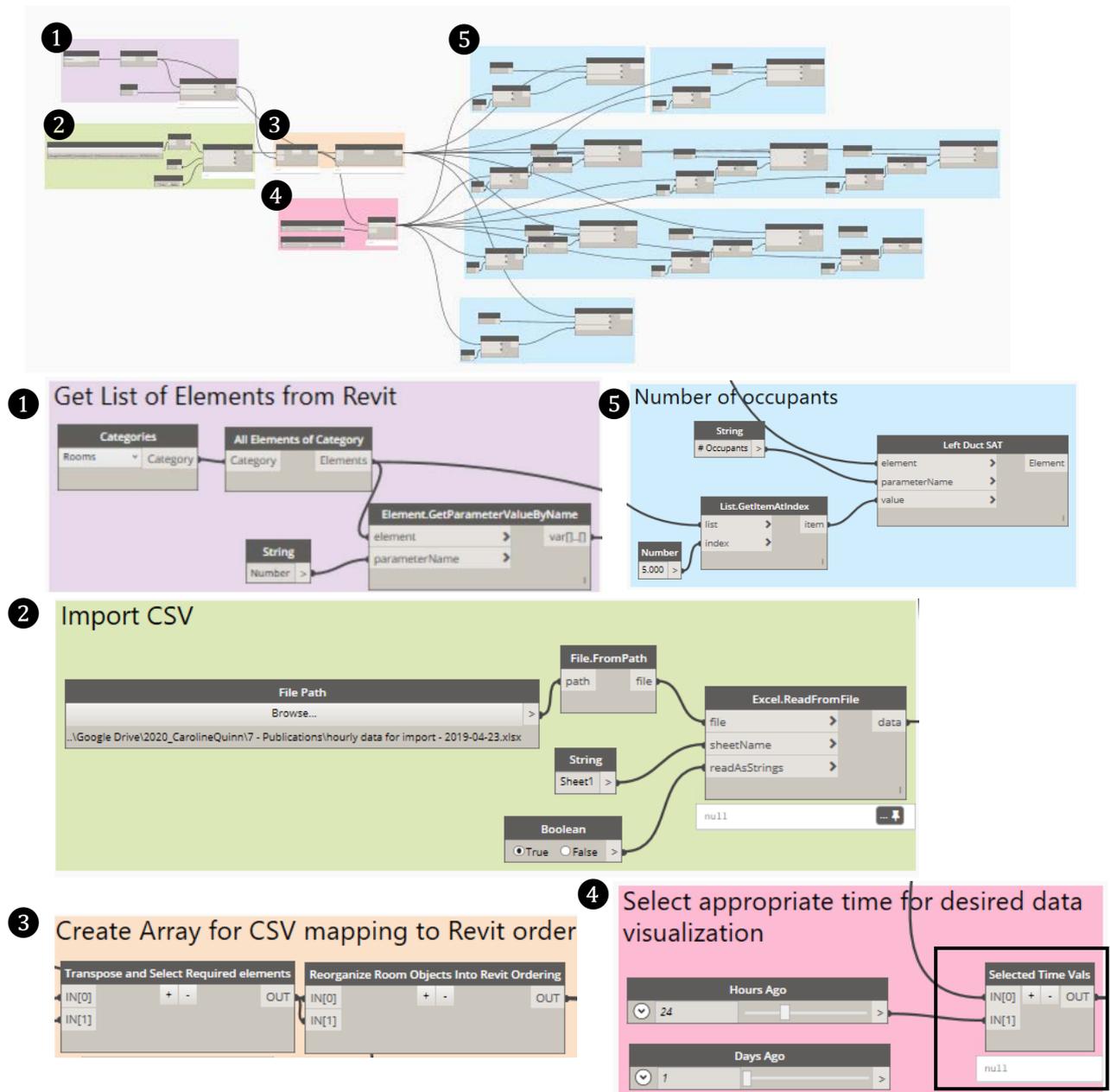

*Figure 14: Dynamo script for time-specific data mapping: high-level summary showing node relationships and enlarged views of individual code blocks*

The 3D list of inputs for mapping has been previously explained in (Bortoluzzi, et al. 2019). The code for selecting the appropriate list in the 3D list (block called out in Figure 14) is as-follows:

# Selected Time Vals

```
import clr

clr.AddReference('ProtoGeometry')

from Autodesk.DesignScript.Geometry import *

allData = IN[0]

selectedTime = IN[1]

valsToMap = [[] for _ in range (len(allData[0][1])-1)]

for j in range (0, len(allData)):

  for k in range (1, len(allData[0][int(selectedTime)])):

    valsToMap[k-1].insert(j , allData[j][int(selectedTime)][k])

OUT = valsToMap
```

Figure 15 shows the five steps of the data transformation process during integration. In step 1, the BIM elements are retrieved from the BIM, this is a table of the element IDs as defined by the BIM and their corresponding BASID which relates to a BASID in the time-series data store. In step 2 the input CSV file with summarized time-series data is imported into the VPL, this data is formatted as previously described. Step 3 creates a temporary 3D list, where a list is created for each data point (POINTID) in the CSV file, i.e. a list for every point relating to the BASID. For example, in the CSV file, ARC.AIR.AHU contains a string of sensor point values related this BASID (an AHU), one of these points is Supply Air Humidity (SAH) and is found in the second place in the value string for any given timestep. The desired time for visualization is selected in Step 4 using time slider; this is shown in a Dynamo node with a sample value of 2. Once the time is selected, the values for each POINTID at the selected time are extracted from the temporary list. In this example the second value, i.e. the average value for each POINT 2 days ago is selected from the temporary 3-D list and stored in a 1-D list. Finally, in step 5, the 1D list is shown as it will be mapped to the BIM fields, where there is a value for each POINTID.

❶ BIM element query

| BIM Element ID | BASID |
|---|---|
| 38526 | ARC.AIR.AHU1 |
| 31429 | ARC.AIR.AHU2 |
| 43512 | ARC.AIR.AHU3 |

❷ Import .csv

| | ARC.AIR.AHU1 | ARC.AIR.AHU2 | ... |
|---|---|---|---|
| 2019-05-24 17:00:00 | 0.23,22.3,... | 0.7,22.0,0... | |
| 2019-05-24 16:00:00 | 0.5,23.3,... | 0.8,25.5,1... | |
| 2019-05-24 15:00:00 | 0.22,23.8,... | 1.1,22.0,1... | |

❸ Temporary 3D list

| Time | ARC.AIR.AHU2.SAH |
|---|---|
| 2019-05-24 17:00:00 | 0.7 |
| 2019-05-24 16:00:00 | 0.8 |
| 2019-05-24 15:00:00 | 23.8 |

| Time | ARC.AIR.AHU1.SAT |
|---|---|
| 2019-05-24 17:00:00 | 22.3 |
| 2019-05-24 16:00:00 | 23.3 |
| 2019-05-24 15:00:00 | 23.8 |

| Time | ARC.AIR.AHU1.SAH |
|---|---|
| 2019-05-24 17:00:00 | 0.23 |
| 2019-05-24 16:00:00 | 0.5 |
| 2019-05-24 15:00:00 | 0.22 |
| ... | ... |

❹ Time Slider

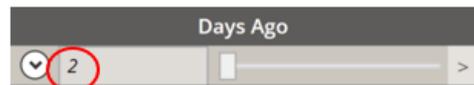

Days Ago

| ⌄ | 2 | | > |

Temporary 3D list

| Time | ARC.AIR.AHU2.SAH |
|---|---|
| 2019-05-24 17:00:00 | 0.7 |
| 2019-05-24 16:00:00 | 0.8 |
| 2019-05-24 15:00:00 | 23.8 |

| Time | ARC.AIR.AHU1.SAT |
|---|---|
| 2019-05-24 17:00:00 | 22.3 |
| 2019-05-24 16:00:00 | 23.3 |
| 2019-05-24 15:00:00 | 23.8 |

| Time | ARC.AIR.AHU1.SAH |
|---|---|
| 2019-05-24 17:00:00 | 0.23 |
| 2019-05-24 16:00:00 | 0.5 |
| 2019-05-24 15:00:00 | 0.22 |
| ... | ... |

❺ 2D List for mapping to BIM

| 0.5 |
|---|
| 23.2 |
| 0.8 |

*Figure 15: Data transformation during integration.*

Note that the scripts presently do not support null point values as this causes data type errors within the Dynamo and therefore a large dummy value (555555) was inserted for missing data points. Null point values can occur when data points use a REpresentational State Transfer (REST) integration transfer protocol, this is true for controllers using the BACnet transfer protocol for example. When a REST transfer protocol is used data point values are only provided on request, and there is therefore not a value for every point at each timestep. Future work is necessary to overcome this issue and permit null value mapping so that sparse matrices of point value data can be handled, as these are computationally the most efficient for data storage.

Using the integration method, data has been mapped to an FM-BIM for the small case study host building created in Revit 2019. Data points relating to ROOMIDs can be directly mapped to the FM-BIM, there BASID data values require a lookup to a spatial table to cross reference the equipment to a location in the building. The Dynamo script presented previously has been integrated with this FM-BIM and a sample visualization of the daily average temperature data is shown in Figure 16. Because all current data is mapped to the BIM, the other summarized sensor data values are also available by viewing the room or element properties, as shown in Figure 17.

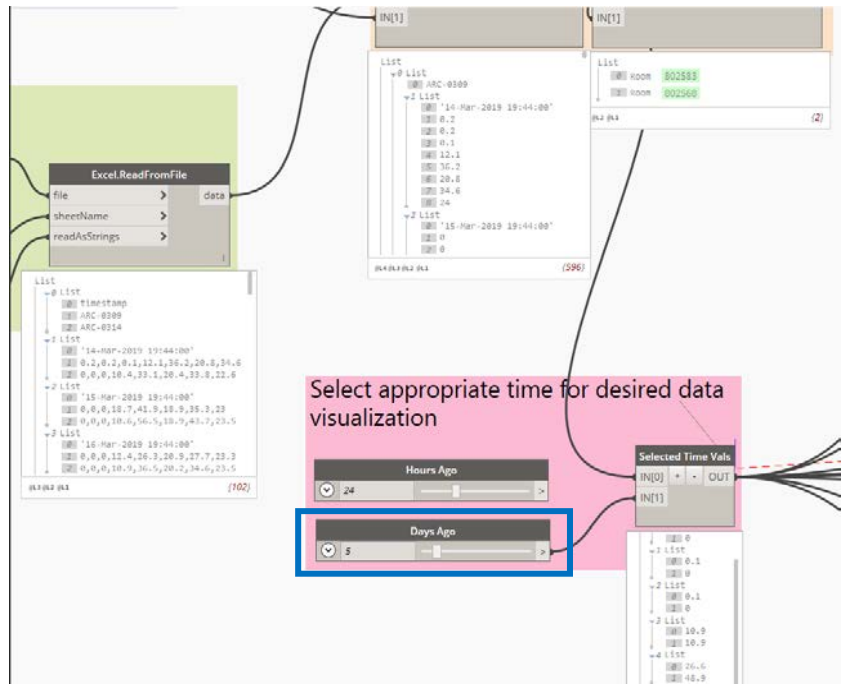

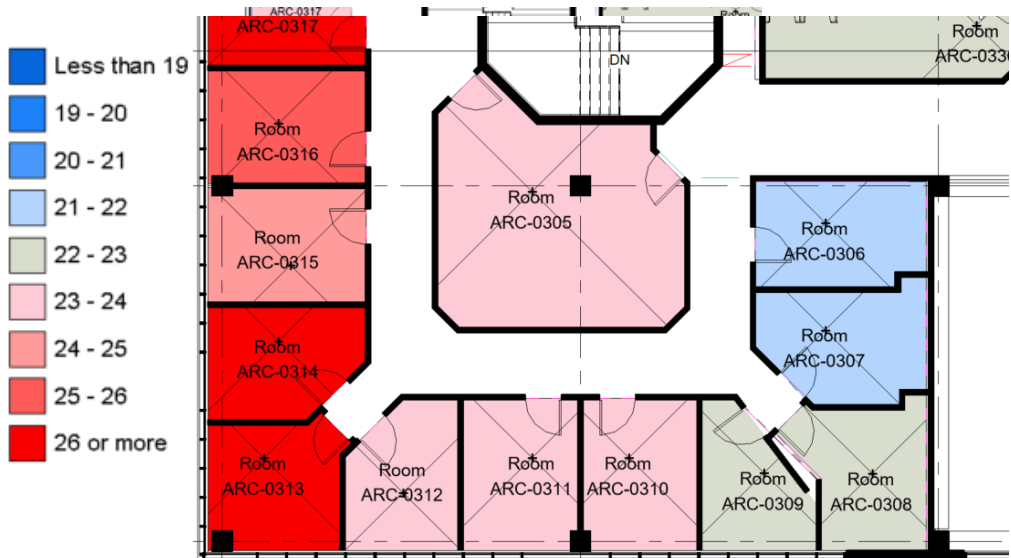

*Figure 16: Top: Inputting desired time in Dynamo "5 days ago". Bottom: FM-BIM visualizing average temperature for "5 days ago"*

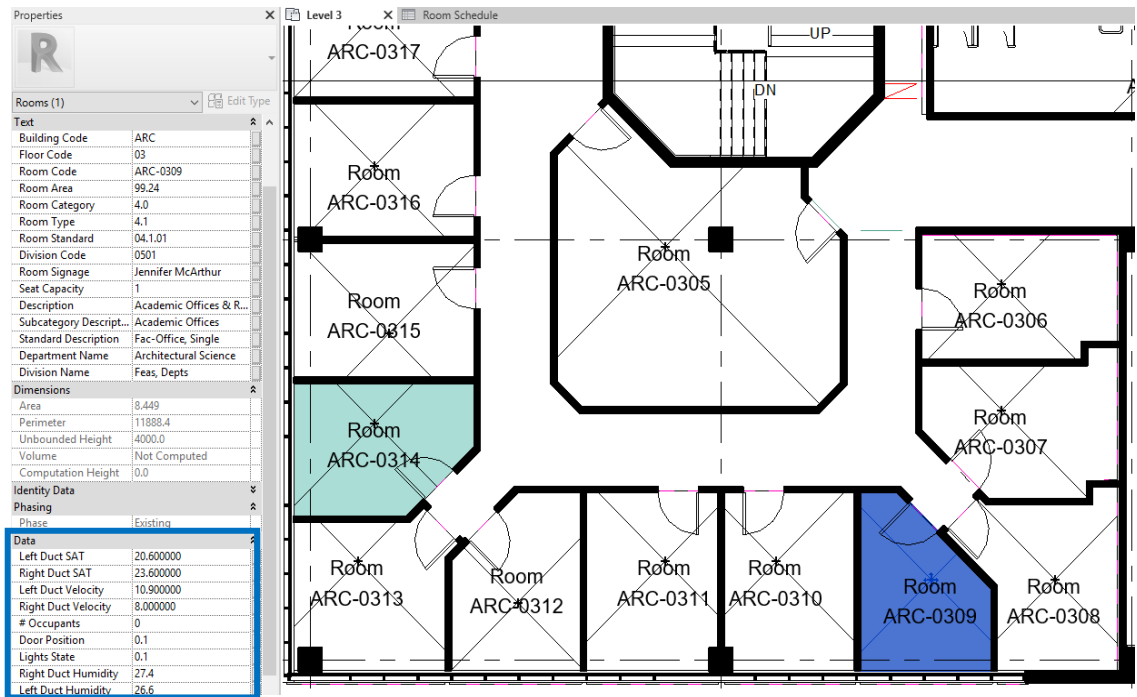

*Figure 17: All sensor data averages as measured 5 days ago*

## 6 Discussion

BAS data is highly valuable within an FM-BIM. However, because BAS point naming conventions, particularly in legacy systems, are frequently a) not human-readable, b) indicate the system and equipment measured but not the network location, or c) indicate the network location of the point but not the equipment or system being measured, it can be extremely difficult to structure using a formal ontology. Further, due to high measurement frequency, the supporting database architecture and data visualization must be carefully considered in order to facilitate BAS to FM-BIM integration.

This paper presents such an architecture that allows for high-frequency low-latency data transmission by using sensor point controllers, dedicated database buffers and streamers, a *directly linked* cloud-based database, and batch analytics. This architecture permits efficient summarization of heterogeneous BAS data on the cloud where scalable computational resources are available. The data summarization techniques proposed include averages and counts on an hourly, daily, and monthly basis. Consistent with building energy modeling norms, data regarding occupancy or lighting state are indicated in real time as integers; however, these must be stored as floats for hourly, daily, and monthly averages to better reflect the percentage time that a space is occupied or illuminated. The summarization of sensor data using this method allows the FM-BIM to be the single building data model for FM users to consult during the operational life of a building, while maintaining a data structure that can also be used by building applications for energy management and other building controls.

Previously *directly linked* approaches for integration between IoT/BAS and BIM have not been optimized for querying and visualization. The proposed approach offers a methodology for time-series data integration and summarization which allows for such querrying and visualization, further enhansed by the support usesr navigation over multiple time granularities. User access to such data ultimatlry allows for data driven decision making. the proposed acquisition and ingestion methodology can further be used in the integration of IoT and BAS data to other Smart Building Applications.

The use of Dynamo, a VPL, was found to be highly effective in mapping time-series data integration from a CSV file to an FM-BIM, provided proper structure. The presented format contains sensor data for each point as a string to form a tuple, which is then inserted into a 2D list with time in rows and BASIDs in columns to create a 3D list with indices of time,

BASID, and point values. From this list, sub-arrays mapping BASID vs POINTID can be accessed for a given time using the slider and code presented. This study has been limited to conceptualizing the larger scale building implementation, currently only a single element type can be mapped at one time, and a more complex mapping algorithm will be required to update both room-hosted and element-hosted time-series data at the same time. An additional limitation of this work requiring further development is that null values cannot be processed by the script; as a temporary solution, a dummy value (555555) has been substituted for missing data, but code refinement is necessary in future work to overcome this limitation. This will reduce the computational cost by permitting the use of sparse matrices for data storage.

Cyber physical systems are concerned with the feedback loop of sensing, evaluating, and acting on building conditions. Schmidta & Åhlund (Schmidta and Åhlund 2018) describe building automation as a three-layer architecture facilitating this feedback loop. This research focuses on making the bottom layer of this architecture- the field level comprised of sensors and actuators - available to the top layer - the management level comprised of building management system and visualizations tools while maintaining the ability to engage the middle layer where applications such as predictive control are applied. Most current research in cyber physical systems focuses on optimizing operational energy cost or consumption (Schmidta and Åhlund 2018) through the development of the middle layer. However, sensor data visualization should be considered and further researched in cyber physical systems. The middle layer is not easily interpreted by FM users, and without a thoroughly developed top management layer, there will be a loss of agency for FM users as cyber physical system development progresses. This is increasingly important for cyber physical systems with tightly coupled sensing and actuating embedded systems as described by Kleissl & Agarwal (2010) where there is no discussion of visualization at the management level.

## 7   Conclusions

The move toward digitized building information has removed some access to building information from FM. An FM-BIM with integrated summarized IoT sensor, actuator, and meter data would reintroduce access to this information, enabling FM to complete more complex analysis than previously possible. Appropriate time granularity of sensor data will be available to FM through the use of time sliders, which map summarized data of common metrics: min/max, averages, and counts to corresponding BIM fields for each datapoint, and are visualized using color coding. A linked data architecture is presented in this research, where Dynamo is used to link IoT data and the BIM. Although functional, Dynamo is not optimized to query databases, a query processor which can directly receive time slider requests and return relevant data would be more efficient. Furthermore, an appropriate query processor would remove the need for a custom CSV file to be linked in Dynamo before using the time slider. While the linked data architecture with predetermined summary tables has been successful in this study, a centralized model with a sophisticated query processor would provide increased flexibility and permit new data summaries to be developed on an as-needed basis. This demonstrates the key limitation of the linked data approach, which is the fixed nature of the architecture.

This research lays the foundation for a long-term project to develop a cloud-hosted BIM-integrated FM platform, permitting data analytics with complex predictive modeling and classification algorithms to support applications such as Smart and Continuous Commissioning and Model Predictive Control. The visualization of the summarized sensor data provides an integrated view for facility managers and building operators to support integrated asset management and optimization. This work has yet to be tested by facility-users and this testing will form the long-term future work in this field. In addition, incremental functionality will be developed to enhance the analytics to integrate deficiency alarms using the trended data as defined in accordance with ASHRAE Guideline 36-2018 High Performance Sequences of Operation for HVAC Systems (ASHRAE 2018). A more refined graphical user interface should be developed, where users can interact with sliders outside of the VPL to initiate IoT sensor data integration with an FM-BIM. Further work on defining relevant BAS data summarization techniques should be done in collaboration with end-users such as Facility Engineers to ensure that the most useful data is presented. Topics for future consideration include additional measures to be mapped, for example, data maxima and minima, different timeframes, and dashboard integration.

## 8    Acknowledgements

This is a substantially extended and enhanced version of the paper presented at the CIB W78 Annual Conference held at Northumbria University in Newcastle UK in September 2019 and incorporates content from the paper submitted to the 2020 conference. We would like to acknowledge the editorial contributions of Professor Bimal Kumar of Northumbria University and Dr Farzad Rahimian of Teesside University in the publication of this paper. This research was financially supported by the Natural Science and Engineering Research Council [CREATE 510284-2018] and CRD [CRDPJ 461929 - 13] programs, the Ontario Research Fund Centres of Excellence program (*Big Data Research, Analytics, and Information Network*), and FuseForward Solutions Group. Finally, this project would not have been possible without the active support of Ryerson's Facility Management and Development team.

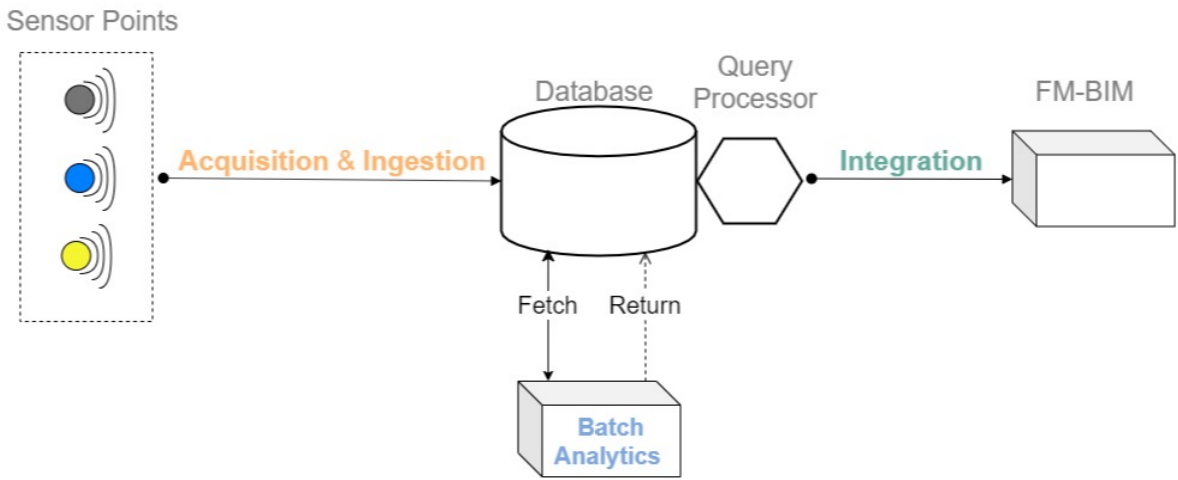

*Figure 1:* System architecture for IoT data streaming to FM-BIM

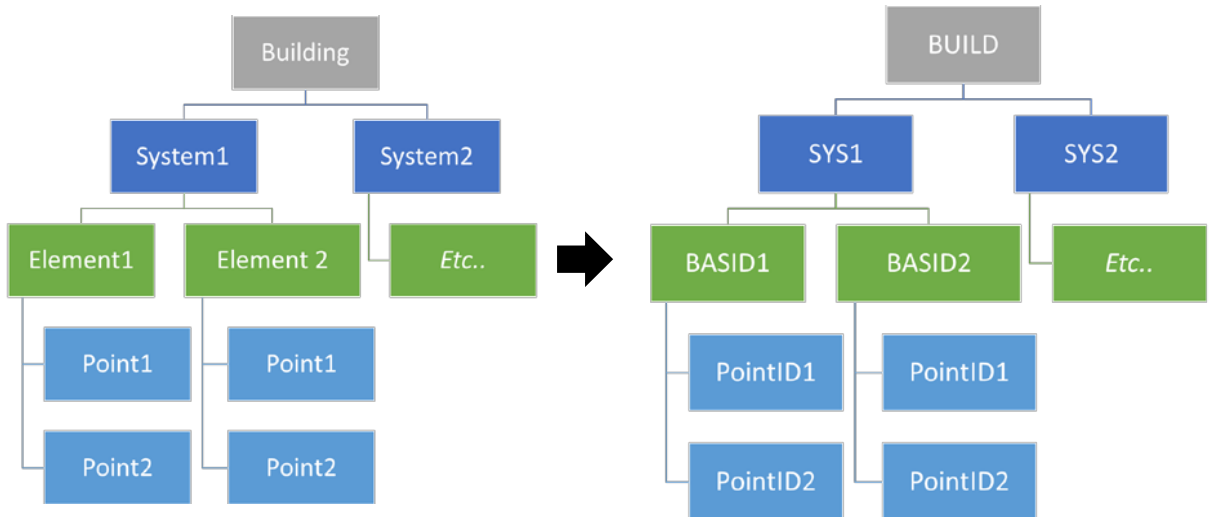

*Figure 2: Database Hierarchical Schema – general (left) and controls-specific*

*implementation in this paper (right)*

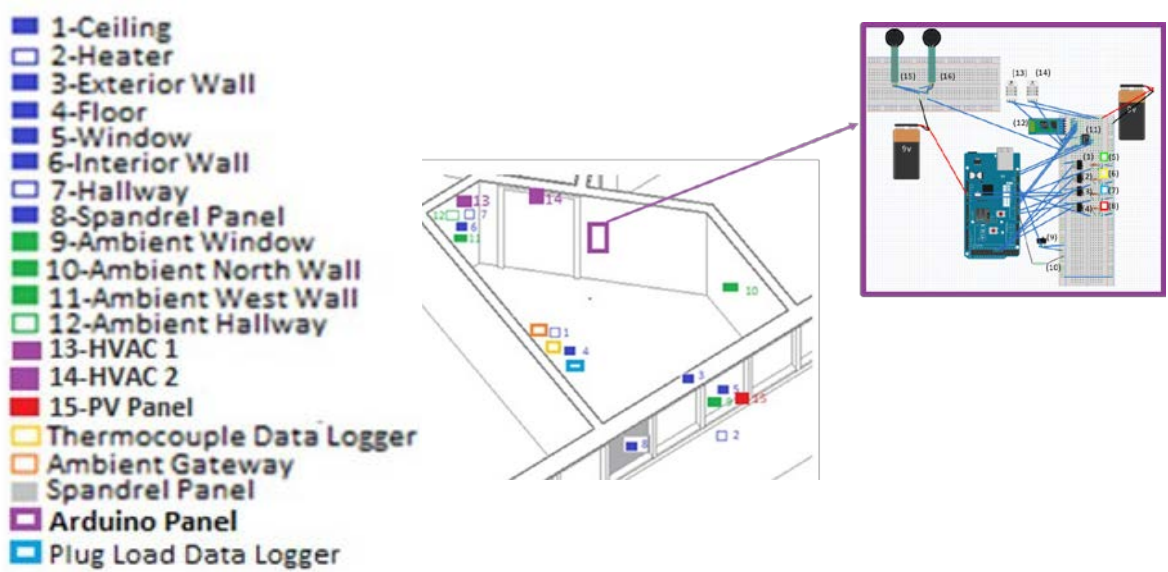

**Legend:**
- 1-Ceiling
- 2-Heater
- 3-Exterior Wall
- 4-Floor
- 5-Window
- 6-Interior Wall
- 7-Hallway
- 8-Spandrel Panel
- 9-Ambient Window
- 10-Ambient North Wall
- 11-Ambient West Wall
- 12-Ambient Hallway
- 13-HVAC 1
- 14-HVAC 2
- 15-PV Panel
- Thermocouple Data Logger
- Ambient Gateway
- Spandrel Panel
- Arduino Panel
- Plug Load Data Logger

| | A | B | C |
|---|---|---|---|
| 1 | timestamp | ARC-0309 | ARC-0314 |
| 2 | '14-Mar-2019 19:44:00' | 0.2,0.2,0.1,12.1,36.2,20.8,34.6,24 | 0,0,0,10.4,33.1,20.4,33.8,22.6 |
| 3 | '15-Mar-2019 19:44:00' | 0,0,0,18.7,41.9,18.9,35.3,23 | 0,0,0,10.6,56.5,18.9,43.7,23.5 |
| 4 | '16-Mar-2019 19:44:00' | 0,0,0,12.4,26.3,20.9,27.7,23.3 | 0,0,0,10.9,36.5,20.2,34.6,23.5 |
| 5 | '17-Mar-2019 19:44:00' | 0,0,0,11,21,21.2,23.8,23.6 | 0,0,0,10.9,45.4,19.7,38.4,23.5 |
| 6 | '18-Mar-2019 19:44:00' | 0.1,0.1,0.1,10.9,26.6,20.6,27.4,23.6 | 0,0,0,10.9,48.9,19.1,40,23.7 |
| 7 | '19-Mar-2019 19:44:00' | 0.3,0.3,0.2,13.4,30.4,20,29.7,23.4 | 0,0,0,11,21,21.2,23.8,23.6 |

*Figure 3: Sample of IoT data streaming (top) office living lab set up (bottom) and pre-processed daily average CSV rows*

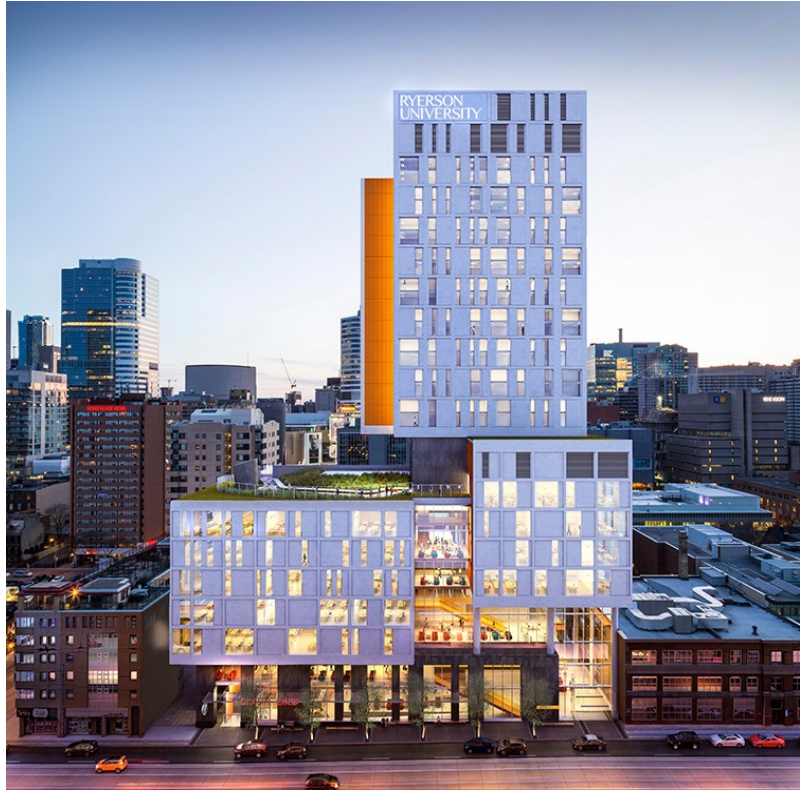

*Figure 4: Daphne Cockwell Complex at Ryerson University* (Perkins + Will 2019)

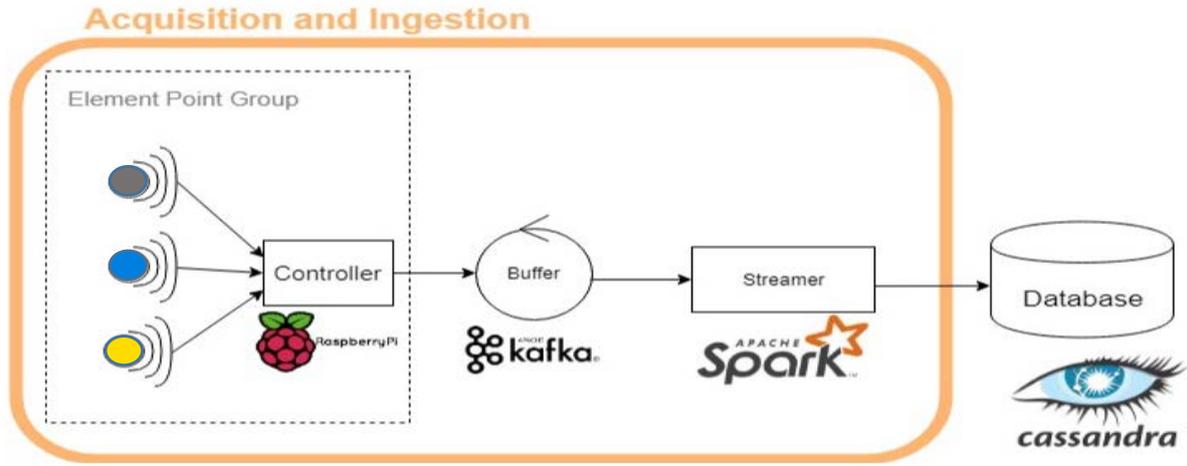

Sensors with unique Identifiers: **BUILD.SYS.BASID.POINTID**

*Figure 5: Data acquisition and ingestion process*

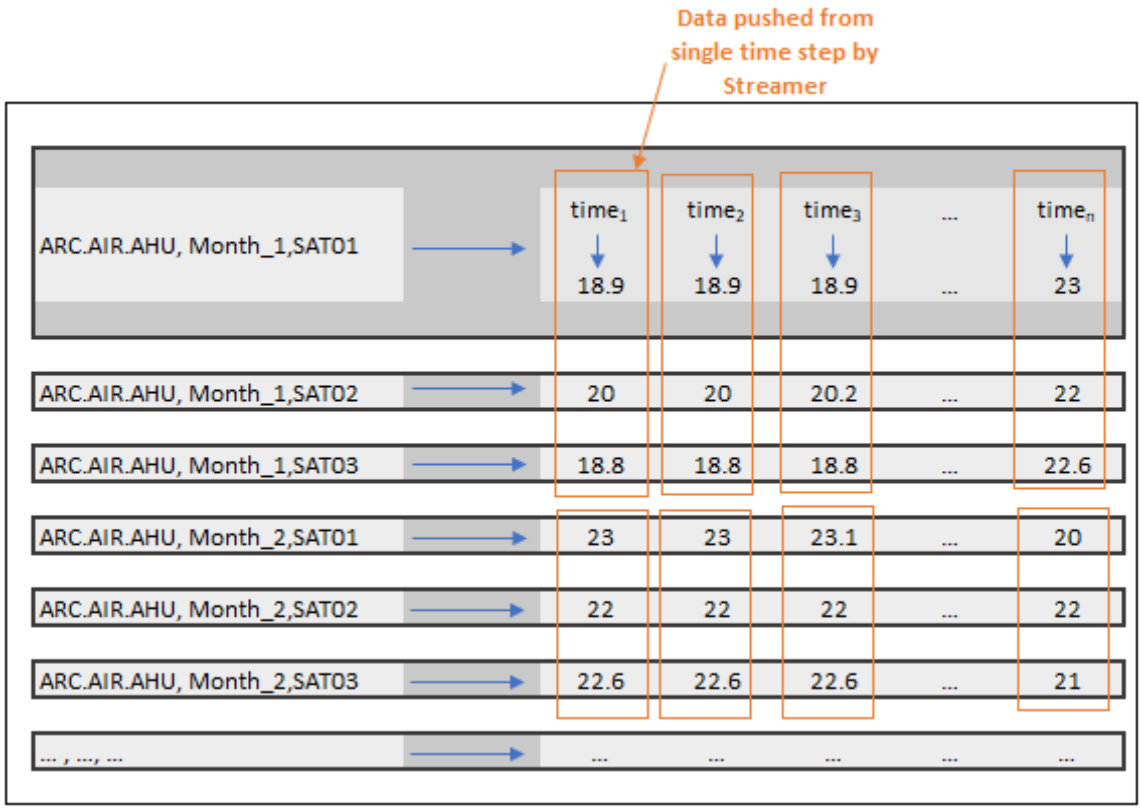

*Figure 6: IoT data storage in Cassandra*

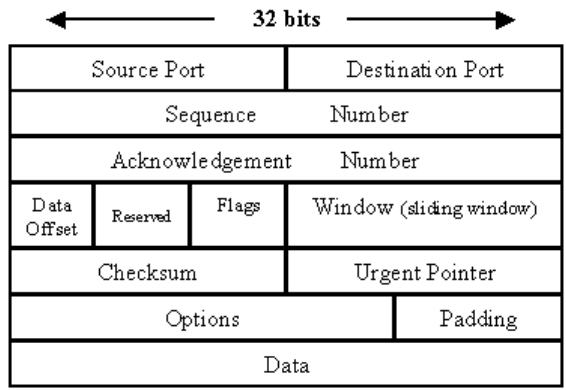

*Figure 7: TCP packet structure* (Salomon 2006)

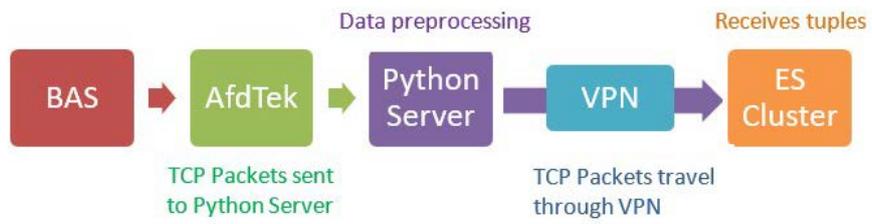

*Figure 8: Schematic of data streaming architecture from BAS to ES Cluster.*

*Figure 9: Window showing data streaming to ES server via VPN*

| Time | asset_name | meter_name | reading_value | _id |
|------|-----------|-----------|--------------|-----|
| January 29th 2020, 00:03:16.000 | DCCNAE-03 | DCCNAE-03/FC-1.CAV-3-16.T | 21.93143 | d2d45c4f1d5355dec20bd0369J6c60546 |
| January 28th 2020, 08:02:12.000 | DCCNCE-20 | DCCNCE-20/FCB.HTN-20.AI-1 | 1009.604 | 83b8895a7d75e4ec82dbe30d46cd3e561 |
| January 29th 2020, 08:01:57.000 | DCCNAE-04 | DCCNAE-04/FC-1.CAV-4-36.Q | 441 | e96043a966c8e30f66029620094c0Gedz |
| January 29th 2020, 08:01:40.000 | DCCNAE-06 | DCCNAE-06/FC-1.CAV-6-42.Q | 514 | 949dc5fddee431e37a4d4f9a6b02933ec |
| January 29th 2020, 08:01:24.000 | DCCNAE-03 | DCCNAE-03/FC-1.CAV-3-17.HC-O | 70.0466 | 60bcc057e5169Dcd4ad5e4cadc9de58c |
| January 29th 2020, 08:00:07.000 | DCCNAE-05 | DCCNAE-05/FC-1.CAV-5-44.Q | 206 | 3be48b419c69a49582d70101876cbea6 |
| January 29th 2020, 07:59:45.000 | DCCNAE-04 | DCCNAE-04/FC-1.CAV-4-27.HC-O | 46.05006 | e921a556ae9f7597370563356nf6868a6 |
| January 29th 2020, 07:59:30.000 | DCCNAE-06 | DCCNAE-06/FC-1.FCU-6-2.Q | 623 | 562ae031cf8b0963Jb049b875b1a196f |
| January 29th 2020, 07:59:15.000 | DCCNAE-03 | DCCNAE-03/FC-1.CAV-3-17.Q | 629 | b163b15f20a2570464e2619cbd6931ea |
| January 29th 2020, 07:58:04.000 | DCCNAE-05 | DCCNAE-05/FC-1.CAV-5-05.Q | 246 | f468b7035ee2a834776a97111c674d0c |
| January 29th 2020, 07:57:48.000 | DCCNAE-04 | DCCNAE-04/FC-1.CAV-4-27.HC-O | 46.51545 | e845da04318bddadcd04ffbd476b35a32 |
| January 29th 2020, 07:57:33.000 | DCCNAE-07 | DCCNAE-07/FC-1.CAV-7-31.T | 22.04713 | d39e6007dfcc3e1e608d62cae47ea7ad |
| January 29th 2020, 07:57:16.000 | DCCNAE-03 | DCCNAE-03/FC-1.CAV-3-39.Q | 671 | 72fcfea908bd0f87fa519fe41c7ea092 |
| January 29th 2020, 07:57:00.000 | DCCNAE-03 | DCCNAE-03/FC-1.CAV-3-14.Q | 437 | e5d014516d5c4bc16dfe58d7d52337b0 |
| January 29th 2020, 07:55:46.000 | DCCNAE-05 | DCCNAE-05/FC-1.CAV-5-44.Q | 202 | 4d49255eef3c6f6e682d61b956060cade |
| January 29th 2020, 07:55:29.000 | DCCNAE-01 | DCCNAE-01/CARMA L1 DACnet IP1.CARMA METER - EMP6.Analog Values.AV-115 | 118.55 | e0c2e0d02c146ccbd9c70d005f24e09e |
| January 28th 2020, 07:55:13.000 | DCCNAE-06 | DCCNAE-06/FC-1.CAV-6-42.Q | 510 | b94bbf188b9b57b53b188a7f18c487Bc |
| January 28th 2020, 07:54:54.000 | DCCNAE-03 | DCCNAE-03/FC-1.CAV-3-14.Q | 442 | 1c7553b8e77d6bf03eb68d56444f2265 |
| January 29th 2020, 07:53:38.000 | DCCNAE-05 | DCCNAE-05/FC-1.CAV-5-05.SAT-SP | 20.77304 | 8daa3305e32635051a7d644e4515b00635 |

*Figure 10: Raw storage of events as recorded on ES Cluster*

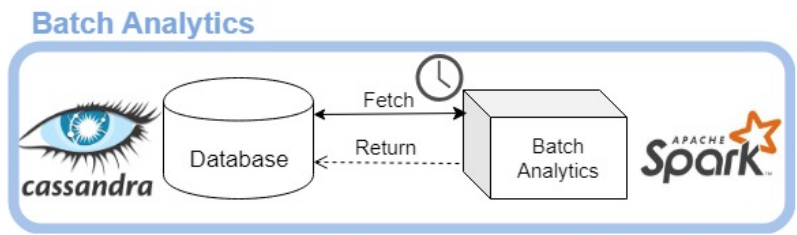

*Figure 11: Batch analytics process*

Time series

Summarized

| Time | ARC.AIR.AHU1.SAT |
|------|------------------|
| 2019-05-02 13:45:50 | 18.9 |
| 2019-05-02 13:45:55 | 18.9 |
| 2019-05-02 13:46:00 | 18.9 |
| 2019-05-02 13:46:05 | 19.0 |
| 2019-05-02 13:46:10 | 19.1 |
| 2019-05-02 13:46:15 | 19.3 |
| 2019-05-02 13:46:20 | 19.5 |
| 2019-05-02 13:46:25 | 19.5 |
| … | … |

| Day | Daily Max ARC.AIR.AHU1.SAT |
|-----|----------------------------|
| 2019-05-02 | 19.5 |
| 2019-05-03 | 23.8 |
| 2019-05-04 | 23.2 |
| 2019-05-05 | 22.4 |
| 2019-05-06 | 26.0 |
| 2019-05-07 | 22.9 |
| 2019-05-08 | 23.5 |
| 2019-05-09 | 19.9 |
| … | … |

*Figure 12: Sample summary of temperature data*

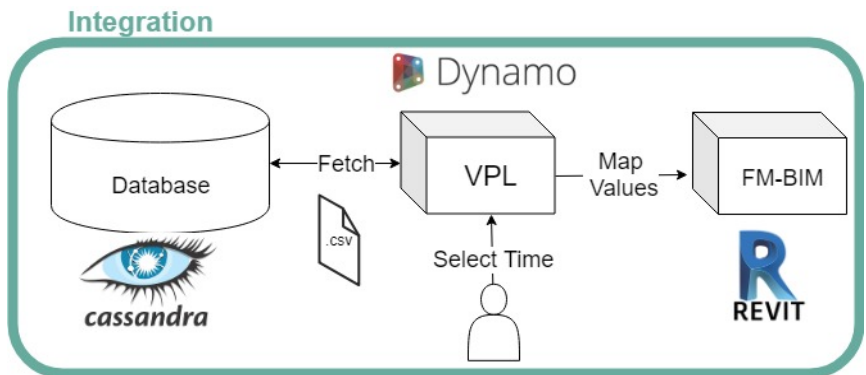

*Figure 13: IoT Sensor Database to FM-BIM integration process*

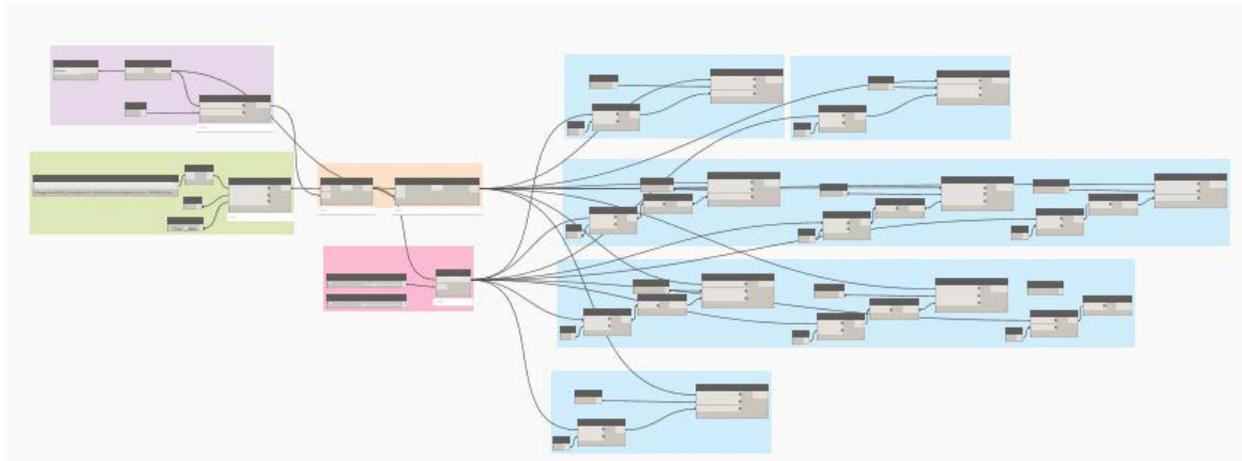

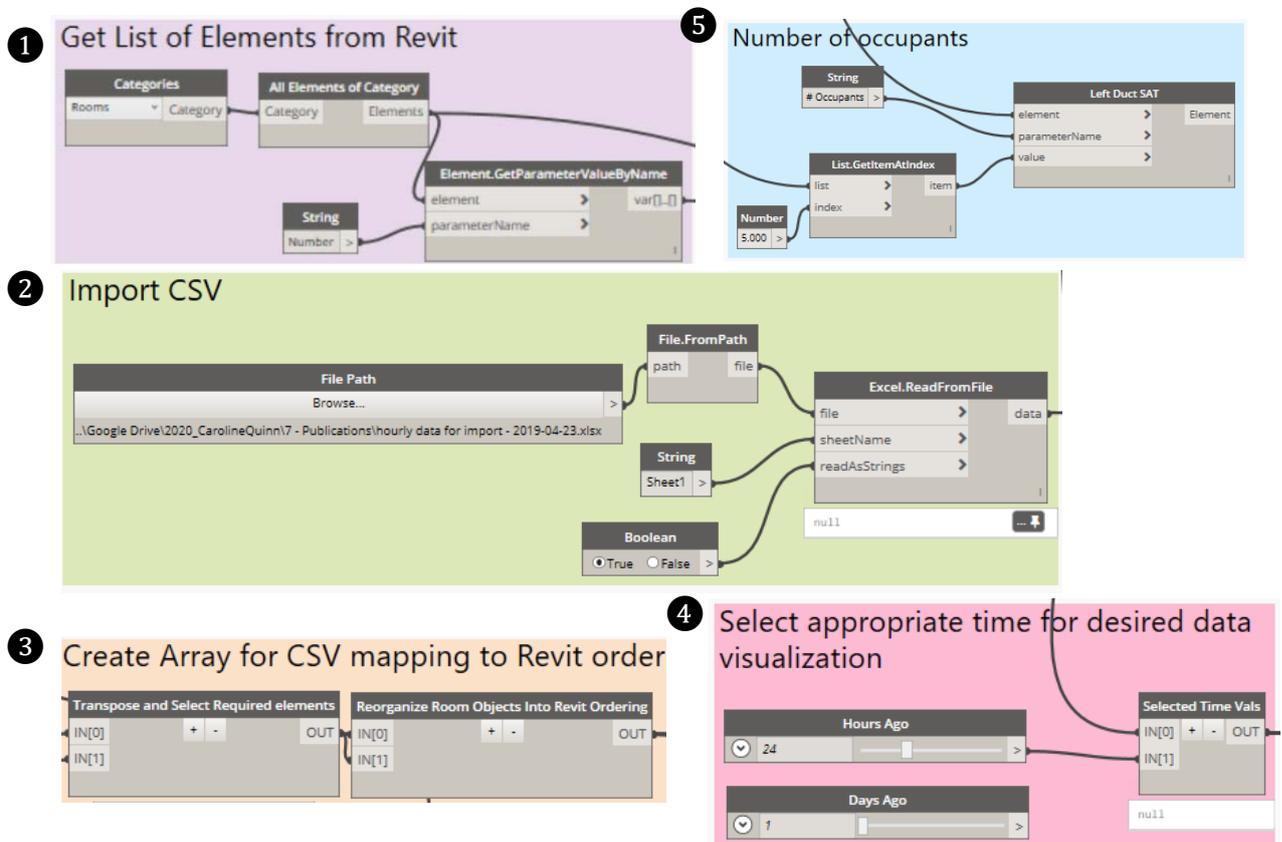

*Figure 14: Dynamo script for time-specific data mapping: high-level summary showing node relationships and enlarged views of individual code blocks*

## ❶ BIM element query

| BIM Element ID | BASID |
|---|---|
| 38526 | ARC.AIR.AHU1 |
| 31429 | ARC.AIR.AHU2 |
| 43512 | ARC.AIR.AHU3 |

## ❷ Import .csv

| | ARC.AIR.AHU1 | ARC.AIR.AHU2 | ... |
|---|---|---|---|
| 2019-05-24 17:00:00 | 0.23,22.3,... | 0.7,22.0,0... | |
| 2019-05-24 16:00:00 | 0.5,23.3,... | 0.8,25.5,1... | |
| 2019-05-24 15:00:00 | 0.22,23.8,... | 1.1,22.0,1... | |

## ❸ Temporary 3D list

| Time | ARC.AIR.AHU2.SAH |
|---|---|
| 2019-05-24 17:00:00 | 0.7 |
| 2019-05-24 16:00:00 | 0.8 |
| 2019-05-24 15:00:00 | 23.8 |

| ... | Time | ARC.AIR.AHU1.SAT |
|---|---|---|
| | 2019-05-24 17:00:00 | 22.3 |
| | 2019-05-24 16:00:00 | 23.3 |
| | 2019-05-24 15:00:00 | 23.8 |

| | ... | Time | ARC.AIR.AHU1.SAH |
|---|---|---|---|
| | | 2019-05-24 17:00:00 | 0.23 |
| | | 2019-05-24 16:00:00 | 0.5 |
| | | 2019-05-24 15:00:00 | 0.22 |
| | | ... | ... |

## ❹ Time Slider

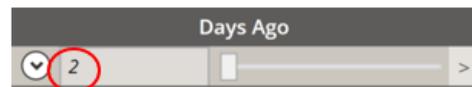

| Days Ago |
|---|

### Temporary 3D list

| Time | ARC.AIR.AHU2.SAH |
|---|---|
| 2019-05-24 17:00:00 | 0.7 |
| → 2019-05-24 16:00:00 | 0.8 |
| 2019-05-24 15:00:00 | 23.8 |

| ... | Time | ARC.AIR.AHU1.SAT |
|---|---|---|
| | 2019-05-24 17:00:00 | 22.3 |
| | → 2019-05-24 16:00:00 | 23.3 |
| | 2019-05-24 15:00:00 | 23.8 |

| | ... | Time | ARC.AIR.AHU1.SAH |
|---|---|---|---|
| | | 2019-05-24 17:00:00 | 0.23 |
| | | → 2019-05-24 16:00:00 | 0.5 |
| | | 2019-05-24 15:00:00 | 0.22 |
| | | ... | ... |

## ❺ 2D List for mapping to BIM

| 0.5 |
|---|
| 23.2 |
| 0.8 |

*Figure 15: Data transformation during integration.*

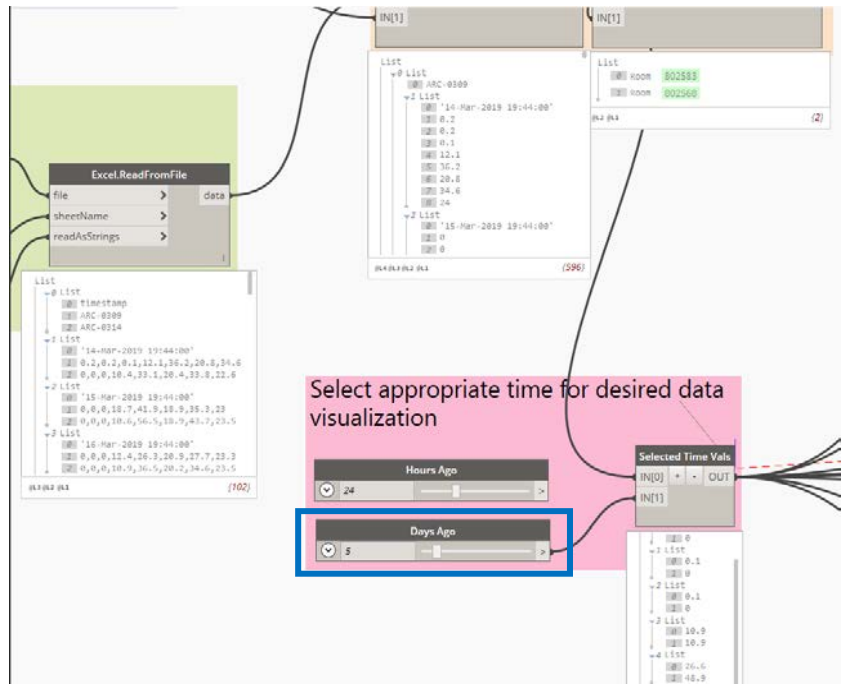

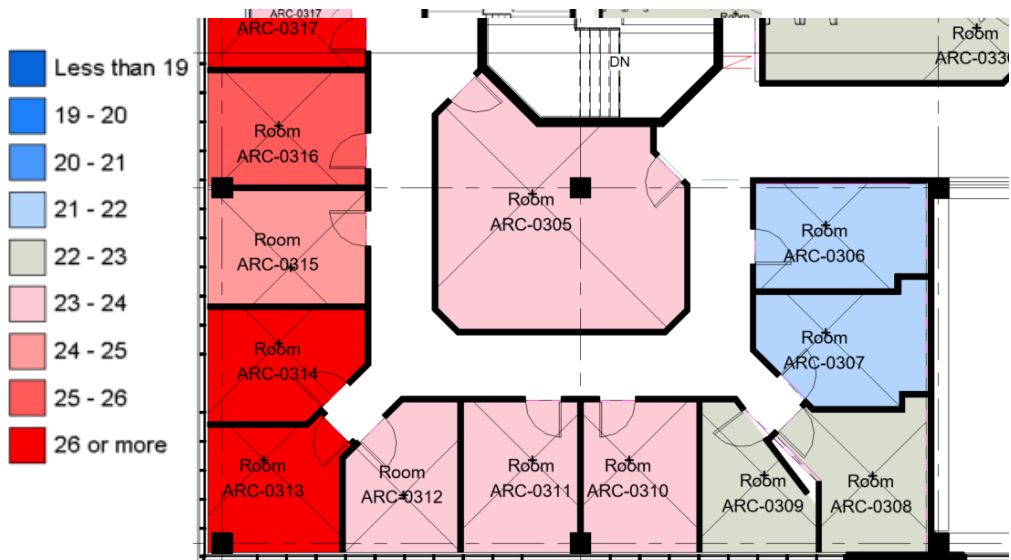

*Figure 16: Top: Inputting desired time in Dynamo "5 days ago". Bottom: FM-BIM*

*visualizing average temperature for "5 days ago"*

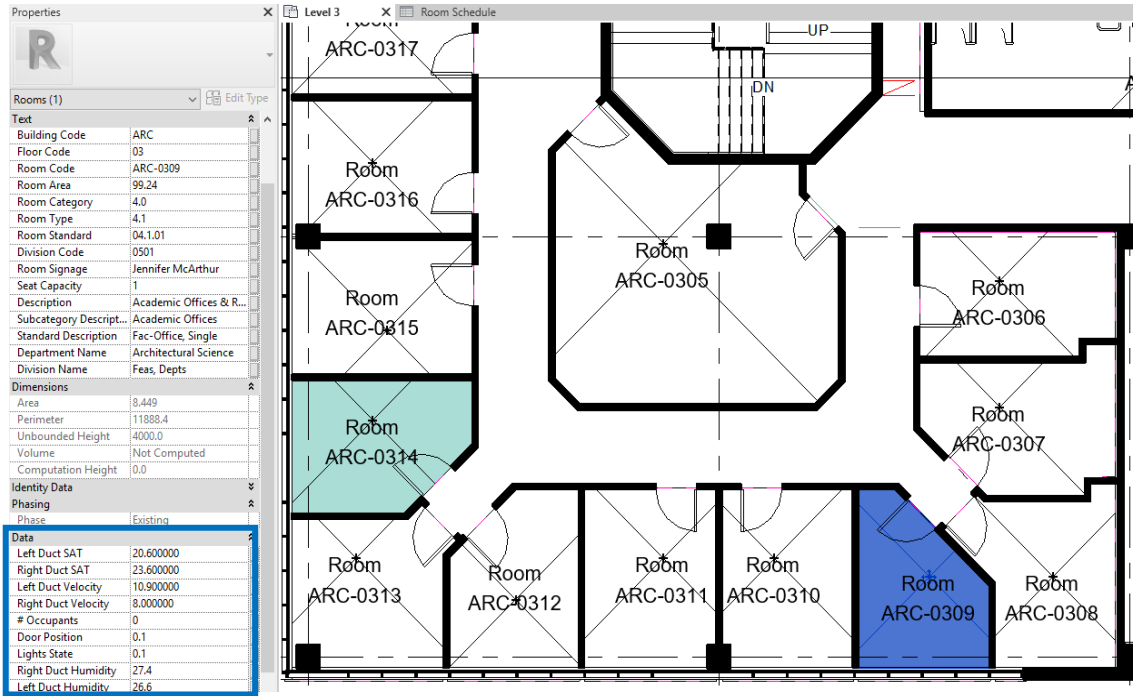

*Figure 17: All sensor data averages as measured 5 days ago*